\documentclass{article}
\usepackage{epsfig} 
\usepackage{eufrak}
\usepackage{overcite}
\DeclareSymbolFont{rsfscript}{OMS}{rsfs}{m}{n}
\DeclareSymbolFontAlphabet{\mathrsfs}{rsfscript}
\newlength{\normalarraycolsep}
\newlength{\normaltabcolsep}
\newcommand{\bm}[1]{\mbox{\boldmath $#1$}}
\newcommand{\bms}[1]{\mbox{\scriptsize \boldmath $#1$}}
\newcommand{\rb}{\bm{r}}
\newcommand{\rbs}{\bms{r}}

\newcommand{\dd}{\mathrm{d}}
\newcommand{\lb}{\ell_B}
\newcommand{\ns}{n_\mathrm{s}}
\newcommand{\nb}{n_\mathrm{b}}
\newcommand{\nc}{n_\mathrm{c}}
\newcommand{\np}{n_\mathrm{p}}
\newcommand{\mup}{\mu_\mathrm{p}}
\newcommand{\mus}{\mu_\mathrm{s}}
\newcommand{\kc}{\kappa_\mathrm{c}}
\newcommand{\kb}{\kappa_\mathrm{b}}

\newcommand{\arctanh}{\mathop\mathrm{arctanh}\nolimits}

\newcommand{\be}{\begin{equation}}
\newcommand{\ee}{\ \ \ \end{equation}}
\newcommand{\bea}{\setlength{\arraycolsep}{0.4\normalarraycolsep}
                  \begin{eqnarray}} 
\newcommand{\eea}{\ \ \ \end{eqnarray}\setlength{\arraycolsep}
                  {\normalarraycolsep}}
\newcommand{\bean}{\setlength{\arraycolsep}{0.4\normalarraycolsep} 
                  \begin{eqnarray*}}
\newcommand{\eean}{\ \ \ \end{eqnarray*}\setlength{\arraycolsep}
                  {\normalarraycolsep}}

\begin{document}
\title{Where the linearized Poisson-Boltzmann 
cell model fails: (I) spurious phase
separation in charged colloidal suspensions}
\author{
M. N. Tamashiro and H. Schiessel\\ 
\textit{Max-Planck-Institut f\"ur Polymerforschung,}\\
\textit{Ackermannweg 10, 55128 Mainz, Germany}
}
\date{}
\maketitle
\begin{abstract}
We perform a linearization of the Poisson-Boltzmann (PB)
density functional for spherical Wigner-Seitz cells that yields  
Debye-H\"uckel-like equations agreeing asymptotically with
the PB results in the weak-coupling (high-temperature) limit. 
Both the canonical (fixed number of 
microions) as well as the semi-grand-canonical (in contact with 
an infinite salt reservoir) cases are considered
and discussed in a unified linearized framework. 
In the canonical case, for sufficiently large colloidal charges 
the linearized theory predicts the occurrence of a 
thermodynamical instability with an associated phase separation
of the homogeneous suspension  
into dilute (gas) and dense (liquid) phases. 
In the semi-grand-canonical case it is predicted 
that the isothermal compressibility and the 
osmotic-pressure difference between the colloidal suspension
and the salt reservoir become negative in the 
low-temperature, high-surface charge
or infinite-dilution (of polyions) limits.
As already pointed out in the literature for the latter 
case, these features are in disagreement 
with the exact nonlinear PB solution inside a Wigner-Seitz cell 
and are thus artifacts of the linearization.
By using explicitly gauge-invariant forms of the electrostatic potential 
we show that these artifacts, 
although thermodynamically consistent with quadratic 
expansions of the nonlinear functional and osmotic pressure, 
may be traced back to the non-fulfillment of the 
underlying assumptions of the linearization.
\end{abstract}

\section{Introduction}

The study of classical charged systems has received 
an increased renewed interest in view of many 
industrial applications \cite{hunter,evans}: paint, 
petrochemicals, food, pharmaceuticals,
cosmetics, diapers, sewage treatment etc. 
Many environmental-friendly new materials
are hydrosoluble due to the presence of ionizable groups that 
dissociate upon contact with water. In fact, their hydrosolubility
is a result of the combination of Coulomb repulsion
between fixed charged monomers and the mixing entropy maximized by 
the mobility in solution of the oppositely-charged small counterions.  
Besides technological applications, charged macromolecules
like lipid aggregates (bilayers, micelles and vesicles), 
proteins and polynucleotides (including DNA and RNA)
are also of fundamental importance in
the biochemistry of living systems \cite{biology1,biology2}.
Furthermore, due to the availability of faster computers, 
many new insights in soft-matter physics come from
Monte-Carlo and molecular-dynamics simulations of charged 
systems~\cite{holm1,holm2}.
These may be partially viewed as controlled 
theoretical experiments and provide a complementary 
approach to analytical treatments. 

An ubiquitous case is that of 
mesoscopic charged colloidal particles
(also called polyions or macroions) immersed in aqueous solution, 
which polarize the small mobile ions in their vicinity:
microions of opposite sign (counterions) are attracted
to them, while like-sign microions (coions) are
repelled. 
The theoretical description of these systems 
requires the understanding of the role of the electrostatic 
interactions between charged objects mediated by the 
surrounding aqueous ionic solution.
In view of the many-body problem and the  
long-range nature of the Coulomb interaction, 
a statistical-mechanical treatment of the system 
is nontrivial.
Within the Primitive Model~\cite{friedman} (PM) the molecular nature 
of the solvent is ignored (neglect of van der Waals and 
hydration forces) and the suspension is treated as a two-component 
system, comprised of the highly-charged large polyions (and its 
neutralizing counterions) and oppositely-charged 
pairs (anions and cations) 
of ionized salt particles. These are immersed 
in a continuous medium of dielectric constant $\epsilon$ 
and interact through the bare Coulomb potential with 
additional hard-sphere repulsion. In the PM it is implicitly assumed 
that the (hard) spheres have the same dielectric 
constant as the solvent, so there are no electrostatic 
image effects. 
For symmetric (in size and charge) electrolytes the PM 
reduces to the Restricted Primitive Model (RPM)
and a theoretical description for dilute 
solutions may be developed using the traditional 
Debye-H\"uckel (DH) theory for 
electrolytes~\cite{debyehuckel,hill,mcquarrie}, with 
some improvements taking nonlinearities~\cite{yanrpm}  
 into account or using integral-equation 
methods~\cite{mcquarrie}.
An extension of these theories for a colloidal suspension 
is nontrivial~\cite{liquidstate1,liquidstate2} in view of the huge 
asymmetry between poly- and microions. Compared to the 
symmetric case, nonlinearities are magnified and dominate 
in the strong asymmetric colloidal limit. 

A mean-field approach to the PM, although not thermodynamically
consistent~\cite{onsager,yancorrelations}, is represented by the 
Poisson-Boltzmann (PB) 
approximation~\cite{israelachvili,safran,desernohouches,netzorland}.
This treatment gives a reasonable description 
in the weak-coupling (high temperature 
or small charge densities) limit, when the
microionic correlations that are neglected at the mean-field
level are unimportant. But even the mean-field 
PB approximation for a suspension of charged 
polyions is a formidable task~\cite{fushiki} due to the 
large asymmetry in size, mass and charge between the polyions and 
the small mobile ions. 
To circumvent this problem, the cell model has been 
introduced --- originally for the cylindrical 
geometry~\cite{fuoss,alfrey} ---  
which reduces the many-body problem 
to the study of a (fixed) single polyion inside a Wigner-Seitz (WS) cell, 
whose volume is related to the polyion density.
In the WS-cell model the single polyion plays only the role of 
a \textit{boundary condition.} Note that this represents a dramatic 
simplification to the original PM formulation, where polyions and 
microions are treated at the same level. 
Even though the applicability of the WS-cell model appears at first sight to
be only justified for an ordered crystalline 
phase, it has also been used to describe the fluid 
phase~\cite{marcus,alexanderpincus}, neglecting thus 
the polyion translational entropy, 
the polyion-polyion interactions and 
\textit{intercell} (both microion-polyion 
and microion-microion) correlations --- 
\textit{intracell} microion-microion correlations are neglected 
because of the mean-field approximation, which computes only the 
\textit{intracell} microion-polyion correlations. 
To simplify further, the geometry of the WS cell is usually taken 
as to match the boundary condition on the polyion 
charged surface. For example, for colloidal particles  
a spherical charged polyion is considered inside a 
concentric spherical WS cell. 
This reduces the problem to the solution of a second-order 
ordinary (rather than partial) differential equation.  
For the spherical geometry this requires the numerical solution 
of the nonlinear PB equation, contrary to the 
planar \cite{lau,behrens,plane} 
 and cylindrical \cite{fuoss,alfrey,tracy} cases, 
when an exact analytical solution is possible.
In analogy to the DH approach to the RPM, 
it would be thus very helpful to formulate a linearized 
version  of the PB approximation for 
WS-cell models.
We should remark, however, that 
the linearized version (at the mean-field level)  
of the WS-cell model does not include any 
\textit{intercell} (neither polyion-microion nor microion-microion) 
correlations and \textit{intracell} microion-microion correlations, 
contrary to the traditional DH approach to the (symmetric) RPM, which 
automatically includes them because 
in the RPM the mean-field contribution --- 
which in the PB WS-cell model comes from the \textit{intracell}
polyion-microion correlations --- vanishes \cite{yancorrelations}. 
Therefore a more appropriate 
interpretation of the linearized equations to be obtained here 
is that they correspond to an expansion about the 
weak-coupling or high-temperature limit of the mean-field equations.

However, expressions obtained within a linearized framework 
should be interpreted with caution, since they may lead to 
artifacts when applied outside their range of validity.
Besides the specific case to be discussed in this paper,  
another example which is clearly an artifact of the linearization 
corresponds to the attractive component to the 
effective interaction between two confined 
colloids induced by charged walls,
predicted under linearized theory~\cite{goulding} 
but in violation to the exact (at mean-field level) 
nonlinear PB repulsion~\cite{neu}.
Although earlier numerical analysis of the nonlinear solution
were in agreement with the linearized theory~\cite{bowen}, 
these were soon ruled out under very general conditions~\cite{neu}.
The disagreement with the rigorous nonlinear results 
might be attributed to flaws in the numerical 
calculations. 
Attempts to include ionic correlations lead indeed to 
attractive contributions to the effective 
interaction~\cite{spalla,yansphere}, but they 
are doubly-screened and thus are not able to overcome 
the repulsive electrostatic 
DLVO \cite{derjaguinlandau,verwey} 
(Derjaguin-Landau-Verwey-Overbeek) component. 
Therefore, experimental evidences of confinement-induced  
attraction~\cite{kepler,carbajal,crocker} and the occurrence of  
metastable superheated crystals~\cite{larsen} can not be 
explained at the PB mean-field level and still remain an open 
question~\cite{grier}.

Almost two decades after the first experimental evidences of 
attraction between like-charged spherical colloids mediated 
by monovalent counterions in bulk deionized aqueous suspensions, 
its existence is still under dispute.
Under the mentioned conditions, electrostatic-stabilized 
colloidal crystals have been investigated by 
Ise \textit{et al.}~\cite{ise}, revealing the presence of 
empty regions (voids) inside the crystal. These experimental 
observations were interpreted as a coexistence between a 
dense crystalline phase and a dilute gas phase. Similar 
voids were also found experimentally in the fluid phase~\cite{ito},
which, in analogy to the critical behaviour of symmetric electrolytes, 
were interpreted as a coexistence between 
dilute (gas) and dense (liquid) fluid phases.
Even fully equilibrated macroscopic 
gas-liquid phase separation has been 
reported~\cite{tata}, although these  experimental
observations have been attributed to the presence 
of ionic impurities~\cite{palberg}.

From the theoretical point-of-view 
attractive interactions between like-charged spheres 
are observed only under special conditions.
For example, they have been seen in Monte Carlo 
simulations in the presence of multivalent 
counterions~\cite{gronbech,linse,allahyarov}
or when the low-temperature ordering of the discrete charges is  
taken into account~\cite{messina}.
Under the conditions described in the previous 
paragraph those controversial experimental findings are either 
attributed to the presence of long-range attractive electrostatic 
interactions between like-charged poly\-ions~\cite{sogami}
or by state-independent volume terms \cite{beresford,chanpre} obtained by 
approximations that involve some kind of 
linearization: 
random-phase approximation \cite{roij1,roij,hansen},
DH pair-distribution functions
augmented by a variational approach 
for the polyion-polyion interactions \cite{warren},
linear-response approximation \cite{denton},
extended DH theory for asymmetric 
electrolytes \cite{chan},  
mean-spherical approximation (MSA) \cite{petris}
and symmetric PB and MSA \cite{bhuiyan}. 
Even though it has been argued by Overbeek
and others~\cite{overbeekrosenfeld}
that the Sogami-Ise attraction~\cite{sogami}
is due to inconsistencies in their thermodynamic 
treatment, the question does not seem to be settled 
yet and discussion is still in progress~\cite{schmitz}. 
This attractive potential is in contrast to the generally
accepted repulsive electrostatic component of the 
DLVO~\cite{derjaguinlandau,verwey} pair potential
between like-charged polyions. However, the purely repulsive nature 
of the polyion-polyion effective pair potential does not 
preclude \textit{a priori} the existence of a 
liquid-gas separation, as has been shown by 
Roij \textit{et al.}~\cite{roij}.
The focus on the polyion-polyion effective interactions 
overlooks the important contribution to the free energy 
due to the polyion-microion interactions.
Because most of the alternative analytical calculations to the 
Sogami-Ise attractive interaction potential requires 
some linearization procedure, the predicted 
liquid-gas coexistence should be analyzed with 
caution. In fact, these predictions disagree 
with simulation results in the presence of (explicit)
\textit{monovalent} counterions \cite{linse,linse1},
when no instabilities have been detected. Moreover, 
there are indications that the observed van der Waals-like 
loops are artifacts due to the linearization, these being 
drastically suppressed  when nonlinearities
are reintroduced in the theory by the use of
renormalized charges~\cite{diehl}.
Furthermore the linearization of the WS-cell semi-grand-canonical 
PB functional --- which describes at the mean-field level the system in 
electrochemical equilibrium with an infinite salt reservoir --- 
yields negative-compressibility, thermodynamically unstable regions 
which are absent 
in a full nonlinear treatment~\cite{grunberg,deserno}.
Although many aspects of these artifacts for the semi-grand-canonical 
case were already reported 
in the literature~\cite{grunberg}, including a general analysis 
of the linearization scheme for various 
geometries~\cite{deserno}, we believe that 
there are still a few subtle points that need to be clarified.  

The purpose of this paper is first to develop a linearization 
scheme suitable to the canonical (fixed amount of microions) case,
by adopting an explicitly gauge-invariant approach. 
For the semi-grand-canonical case, 
it has been shown by Deserno and von Gr\"unberg \cite{deserno} 
that the occurrence of unstable
linearized equations of state depends on the way the linearization 
scheme is performed and on the osmotic-pressure 
definition. By extending our gauge-invariant approach to the 
semi-grand-canonical ensemble, we try additionally to 
shed some light on this question. 
We argue that thermodynamic stability and consistency are in fact 
independent concepts. The gauge-invariant forms of the 
equations of state allow to establish an explicit
correspondence between their nonlinear and linearized versions. 
The linearized equations, although thermodynamically 
self-consistent with quadratic expansions of the nonlinear ones, 
lead to artifacts when their 
underlying assumptions are not satisfied. 
We will show, by using gauge-invariant forms for the 
electrostatic potential, that there is a \textit{unique} linearization
(about the state-independent zero-th order Donnan densities)
that corresponds to the minimization
of the associated linearized semi-grand-canonical functional,
which is also asymptotically exact (at the mean-field level) 
in the weak-coupling (high-temperature) limit. 
Therefore the expansion of the nonlinear functional 
about the state-independent Donnan densities, originally proposed for the 
spherical geometry by von Gr\"unberg \textit{et al.} \cite{grunberg}
--- and generalized for other geometries 
with analogous high symmetry in Ref.~[\citen{deserno}] ---
 is not only
``optimal'', but it is asymptotically exact   
in the weak-coupling limit. 
In a companion paper\cite{plane} explicit analytical comparison
is performed for the planar case, where the exact nonlinear 
solution (at the mean-field level) can be obtained.

The remainder of the paper is organized as follows. 
In Section~\ref{sec:2} the salt-free model
is introduced and the associated nonlinear equations are reviewed. 
In Section~\ref{sec:3} the linearization of the appropriate 
functional is performed, considering three distinct 
physical situations: the salt-free (in the presence 
of neutralizing counterions only) system introduced in 
Section~\ref{sec:2}, 
with fixed amount of added monovalent 
salt (canonical ensemble) and 
in electrochemical equilibrium with an 
infinite monovalent salt 
reservoir (semi-grand-canonical ensemble). 
In Section~\ref{sec:4} we discuss our results in 
comparison to previous works \cite{grunberg,deserno} and 
present some concluding remarks in Section~\ref{sec:5}.
Appendix~\ref{app:a} presents the boundary-density
theorem (at the nonlinear
mean-field level) for the salt-free simplest case.
In Appendix~\ref{app:b} it is shown 
that the linearized Helmholtz free energy 
may be obtained by a Debye charging process 
of the linearized electrostatic energy. 
Appendix~\ref{app:c}~presents the formal 
derivation of the linearized osmotic pressure
for the salt-free case, showing that it corresponds
to a quadratic expansion of 
the corresponding nonlinear osmotic pressure.
In Appendices~\ref{app:d}~and~\ref{app:e} it 
is shown that 
the linearized osmotic pressure in the presence of salt 
for the canonical and 
semi-grand-canonical ensembles, respectively,
are given by the quadratic expansions of 
the corresponding nonlinear osmotic pressures.
In Appendix~\ref{app:f}~the 
self-consistent linearized averaged densities 
for the semi-grand-canonical ensemble are 
obtained by the minimization of the appropriate
functional. 
In Appendix~\ref{app:g}~we compare the linearized 
semi-grand-canonical osmotic pressures obtained by 
different schemes 
of the Legendre transformation connecting the canonical
and semi-grand-canonical ensembles.

\section{Definition of the model\label{sec:2}}

Although the derivation of the PB equation from a free-energy
density functional can be found 
elsewhere~\cite{trizachansen,netzorland,desernohouches},
to introduce the notation and to stress the advantages of an
explicitly gauge-invariant approach, it is essential to rederive it 
in the following. For simplicity, in this section only the 
salt-free case (in the presence of neutralizing counterions only)
is presented. Generalization of the model including
monovalent salt is straightforward and 
briefly described in Appendices~\ref{app:d}~and~\ref{app:e}
for the canonical and semi-grand-canonical cases, respectively. 

The system to be considered is a hard charged sphere 
(spherical polyion) 
of radius $a$ and total charge $-Zq$ distributed uniformly on its 
surface inside a
concentric spherical WS cell of radius $R>a$, where $q>0$ is the 
elementary charge and $Z\gg 1$ is the 
polyion valence. 
The radius $R$ of the WS cell is related to the polyion density
$\np$ such that the total volume of the suspension is equally 
distributed between the polyions, i.e. $\np\equiv
\left({4\pi} R^3/3\right)^{-1},$ whose hard cores occupy a 
fraction $\phi\equiv\left(a/R\right)^3$ of the total volume.     
To ensure the overall WS-cell charge neutrality, there are
$Z$ mobile (positive) point-like counterions, each carrying a 
charge $+q$, that are allowed to move in 
the spherical shell $a< |\rb|\leq R$, whose volume reads  
\be
V=\int\limits_{a<|\rbs|\leq R}\dd^3\rb
=\frac{4\pi}3 \left(R^3-a^3\right) .
\ee
Henceforth, to simplify the notation, 
it will be implicit that all integrations
are performed over the free volume $V$ unoccupied by the 
polyion core --- but also
including the charged surface at $|\rb|=a$ ---  unless otherwise stated.
At the mean-field PB level the counterions are treated as an
inhomogeneous ideal gas and are described by their (continuous) 
 average local number density 
 $n(\rb)\equiv \left\langle\!\!\left\langle\sum_{i=1}^Z 
\delta^3\left(\rb-\rb_i\right)\right\rangle\!\!\right\rangle$,
where  $\delta^3$ is the three-dimensional Dirac delta unction
and the double brackets denote an ensemble (Boltzmann-weighted) 
 average over all positions of the counterions. 
The total charge number density (counterions plus the negative 
surface charge on the sphere), 
\be
\rho(\rb)=n(\rb)-\frac{Z}{4\pi a^2}\delta^3(|\rb|-a),
\label{eqn:rho}
\ee
is related to the reduced electrostatic potential
$\psi(\rb)\equiv\beta q \Psi(\rb)$
by the (exact) Poisson equation,
\be
\nabla^2 \psi(\rb) = -4\pi\lb \rho(\rb), \label{eqn:poisson}
\ee
where $\lb=\beta q^2/\epsilon$ is the Bjerrum length and 
$\beta^{-1}=k_BT$ is the thermal energy at temperature $T$. 
It is implicitly assumed 
that the solvent dielectric constant $\epsilon$ remains the same 
inside the sphere, so image-charge effects due to dielectric 
contrast are absent. 
The formal solution to the Poisson equation~(\ref{eqn:poisson}) may be 
written in terms of the Green function 
 $G_3(\rb,\rb')$,
\be
\psi(\rb)=\lb\int\dd^3\rb'\,G_3(\rb,\rb')\,\rho(\rb') ,
\qquad  \nabla^2 G_3(\rb,\rb') = -4\pi\delta^3(\rb-\rb'),
\label{eqn:pbpsi}
\ee
which in turn allows us to express the mean-field
Helmholtz free-energy functional ${\mathrsfs F}[n(\rb)]$
of a single WS cell as 
\be
\beta{\mathrsfs F}[n(\rb)] = 
\int\dd^3\rb\,n(\rb)\left\{\ln
  \left[{n(\rb)}\zeta^3\right]-1\right\} +
\frac{\lb}2\int\dd^3\rb\,\dd^3\rb'\,
\rho(\rb)\,G_3(\rb,\rb')\,\rho(\rb'),  \label{eqn:pbfunct}
\ee
where $\zeta$ is the thermal de Broglie wavelength of the counterions.
It should be remarked that the mean-field Helmholtz free-energy  
functional~(\ref{eqn:pbfunct}) can be derived from the underlying PM 
Hamiltonian in different ways:
 as the saddle point of the action in a field-theoretic
approach \cite{netzorland}, from a Gibbs-Bogoljubov
inequality applied to a trial product state that decouples 
the original Hamiltonian \cite{desernohouches} or
from a first-cumulant expansion of the density 
functional reformulation of the associated partition
function \cite{lowen}.
The first term of~(\ref{eqn:pbfunct}) represents the configurational entropy 
of the inhomogeneous ideal gas of counterions, 
while the second term corresponds to the electrostatic
energy, which may be rewritten as 
\bea
\beta U &=& 
\frac{\lb}2\int\dd^3\rb\,\dd^3\rb'\,
\rho(\rb)\,G_3(\rb,\rb')\,\rho(\rb')
=\frac12\int\dd^3\rb\,\psi(\rb)\rho(\rb) 
\nonumber\\&=&
\frac1{8\pi\lb}\int\dd^3\rb\,\left[\nabla\psi(\rb)\right]^2
-\frac1{8\pi\lb}\oint_{\partial V} \psi(\rb)\,
\nabla\psi(\rb)\cdot \dd\bm{S} .\label{eqn:elec_energy}
\eea
The surface contributions to the electrostatic energy
 --- the last term of~(\ref{eqn:elec_energy}), 
performed over the boundary $\partial V$ of the free volume $V$ 
 --- vanish due to Gauss' law and 
the overall WS-cell charge neutrality, 
\be
\int\dd^3\rb\,\rho(\rb)=0, \qquad\qquad\mbox{ or }
\qquad\qquad  \int\dd^3\rb\,n(\rb)=Z. \label{eqn:chargeneut}
\ee

The equilibrium counterion-density profile is obtained by minimizing the 
mean-field functional ${\mathrsfs F}[n(\rb)]$ with respect to $n(\rb)$
under the constraint of overall WS-cell charge 
neutrality~(\ref{eqn:chargeneut}).
For this purpose it is convenient to introduce a
translationally invariant (independent of $\rb$) 
Lagrange multiplier $\beta\mu_\mathrm{el}$ 
and to define the extended Helmholtz free-energy functional including 
a Lagrange-multiplier term, 
\be
{\mathit{\tilde\Omega}}[n(\rb)]
\equiv
{\mathrsfs F} -
\mu_\mathrm{el}\int\dd^3\rb\,\rho(\rb)
=
{\mathrsfs F} - \mu_\mathrm{el}\left[\int\dd^3\rb\,n(\rb)-Z\right],
\ee
which is the \textit{analogue} of 
the conjugated semi-grand-canonical functional
$\mathit{\Omega}[n(\rb)]$ 
in the case of non-fixed number of 
counterions (but fixed number of polyions). 
We should not confuse the \textit{virtual} Lagrange multiplier 
$\mu_\mathrm{el}$ with the \textit{a priori} fixed  
chemical potential of counterions $\mu$ in 
the presence of an infinite counterion reservoir, when the natural 
choice for the thermodynamic ensemble is 
the semi-grand-canonical one, 
\be
{\mathit{\Omega}}
\equiv {\mathrsfs F} - \mu\int\dd^3\rb\,n(\rb) , 
\qquad\qquad  \int\dd^3\rb\,n(\rb)\not\equiv Z.
\ee
The Helmholtz free-energy functional ${\mathrsfs F}$ supplemented by 
the Lagrange-multiplier term,
${\mathit{\tilde\Omega}}\neq{\mathit{\Omega}}$, 
\textit{does not} correspond to the semi-grand-canonical 
functional ${\mathit{\Omega}}$, 
because ${\mathit{\tilde\Omega}}$ and ${\mathit{\Omega}}$  
are obtained under the
constraints of a fixed number of counterions $Z$, 
and fixed chemical potential of counterions $\mu$, respectively.
Although this distinction may be seem rather academic, 
it is important to stress that overall charge neutrality 
(ensured by the Lagrange multiplier $\mu_\mathrm{el}$)
and electrochemical equilibrium (imposed by a fixed 
chemical potential of counterions $\mu$) 
are \textit{independent} constraints. 
This prevents misinterpretations  
when treating the Donnan equilibrium 
(the system in electrochemical equilibrium with an infinite salt reservoir),
when both constraints must be simultaneously satisfied.  

Functional minimization of $\beta{\mathit{\tilde\Omega}}$ with respect to
$n(\rb)$ leads to the Euler-Lagrange or stationary 
condition \cite{footnote2}, 
\be
\left.\frac{\delta \beta{\mathit{\tilde\Omega}}}{\delta n(\rb)}
\right|_{n(\rbs)=\bar{n}(\rbs)} = 
\bar{\psi}(\rb) + \ln\left[{\bar{n}(\rb)}\zeta^3\right]
-\beta\bar\mu_\mathrm{el}=0,
\ee  
which yields the  equilibrium counterion 
profile, $\bar{n}(\rb)=
\mathrm{e}^{\beta\bar\mu_\mathrm{el}-\bar\psi(\rbs)}/\zeta^3$.
In the above formulas the equilibrium electrostatic potential, 
$\bar\psi(\rb)$, is obtained by inserting the 
equilibrium counterion profile $\bar{n}(\rb)$ into 
$\psi(\rb)$, Eq.~(\ref{eqn:pbpsi}).
The Boltzmann-weighted equilibrium counterion profile $\bar{n}(\rb)$,
together with the Poisson equation~(\ref{eqn:poisson}), 
yields the nonlinear PB equation, 
\be
\nabla^2\bar\psi(\rb)=-\frac{4\pi\lb}{\zeta^3}
\mathrm{e}^{\beta\bar\mu_\mathrm{el}-
\bar\psi(\rbs)}+\frac{Z\lb}{a^2}\delta^3(|\rb|-a),
\ee
where the equilibrium
Lagrange multiplier,
\be 
\beta\bar\mu_\mathrm{el}=
\ln\left(\nc\zeta^3\right)-
\ln\left\langle\mathrm{e}^{-\bar\psi({\rbs})}\right\rangle, 
\label{eqn:muel}
\ee
is obtained by  imposing the charge neutrality 
constraint~(\ref{eqn:chargeneut}).
We introduced the \textit{effective} average density 
of counterions in the free volume $V$ unoccupied by the polyion core,
\be
\nc \equiv \left\langle \bar{n}(\rb) \right\rangle = \frac{Z}{V},
\label{eqn:uniformdensity}
\ee
the brackets denoting unweighted spatial averages over $V$, 
\be
\left\langle {\cal X}(\rb)\right\rangle\equiv
\frac{\int\dd^3\rb\, {\cal X}(\rb)}{\int\dd^3\rb} ,
\ee
in contrast to their \textit{nominal} mean density in 
the suspension, $\tilde\nc\equiv \nc\left(1-\phi\right)$, 
which does not take into account the polyion hard cores.
Note that the Lagrange multiplier~(\ref{eqn:muel}) may be 
decomposed into two terms, a chemical potential corresponding 
to an ideal gas of uniform density $\nc$
and an electrostatic contribution due to the counterion-cloud 
polarization, which may be written in terms of an average related 
to the equilibrium 
electrostatic potential $\bar\psi({\rb})$. 
We note that the presence of the Lagrange multiplier $\bar\mu_\mathrm{el}$
ensures \textit{explicitly} the gauge invariance of the equations, that is, 
physical observables, like the equilibrium counterion-density profile,
\be
\bar{n}(\rb) = 
\frac{Z \mathrm{e}^{-\bar\psi(\rbs)}}
{\int\dd^3\rb\,\mathrm{e}^{-\bar\psi(\rbs)}}
=
\frac{\nc\exp\left[{\left\langle\bar\psi\right\rangle-\bar\psi(\rb)}\right]}
{\left\langle\exp\left[{\left\langle\bar\psi\right\rangle-\bar\psi(\rb)}
\right]\right\rangle}, \label{eqn:pbprofile}
\ee
and the electric field, $\bm{E}(\rb)=-\nabla\Psi(\rb)=
-(\beta q)^{-1}\nabla \bar\psi(\rb)$, 
\textit{clearly} do not depend on a particular
choice of the zero of the electrostatic potential $\bar\psi({\rb})$,
since observables can always be written in terms of the gauge-invariant
difference $\beta\bar\mu_\mathrm{el}-\bar\psi(\rb)$.
In particular, explicitly gauge-invariant forms of the 
density profiles will be \textit{useful} to establish
a connection between the nonlinear and the linearized
osmotic pressures~(cf.~Appendices~\ref{app:c}~to~\ref{app:e})
and to derive the self-consistent 
linearized averaged densities for the semi-grand-canonical 
ensemble~(cf.~Appendix~\ref{app:f}). 
We should remark that --- because consistent theories should always  
be gauge invariant --- the use of \textit{explicitly} 
gauge-invariant forms constitutes just a technical convenience,
it does not represent an essential requirement. However, 
we believe that it provides a systematic
and more transparent way to perform the calculations. 
Henceforth \textit{gauge-invariant} will be a short-writing 
for \textit{explicitly gauge-invariant,} unless otherwise stated.

The mean-field Helmholtz free energy (of a single WS cell) $F$ 
is obtained by inserting the equilibrium
density profile $\bar{n}(\rb)$, Eq.~(\ref{eqn:pbprofile}), into
the mean-field functional ${\mathrsfs F}$, Eq.~(\ref{eqn:pbfunct}),  
\bea
\beta F&\equiv&\beta{\mathrsfs F}[\bar{n}(\rb)]= 
\frac12\int\dd^3\rb\,\bar\psi(\rb) \bar{n}(\rb)
-\frac12 Z\bar\psi(a)
+ \int\dd^3\rb\,\bar{n}(\rb)\left\{\ln \left[{\bar{n}(\rb)} \zeta^3\right]-1\right\}
\nonumber\\
&=& Z\left[\beta\bar\mu_\mathrm{el}-1-\frac12 \bar\psi(a)\right]
-\frac12\int\dd^3\rb \,\bar\psi(\rb)\bar{n} (\rb).\label{eqn:pb}
\eea
In Appendix~\ref{app:a} it is shown that the 
(nonlinear) osmotic pressure $P$ (over pure solvent), 
defined as the negative \textit{total derivative} of the Helmholtz
{free energy} $F$, Eq.~(\ref{eqn:pb}), 
with respect to the WS-cell free volume \cite{footnote3} $V$, 
is simply given by 
\bea
\beta P &\equiv& \left.-\frac{\dd \beta F}{\dd V}\right|_{Z,T}= \bar{n}(R),
\label{eqn:bounddensity}
\eea
which is the well-know WS-cell mean-field result that the salt-free
osmotic pressure is related to the counterion boundary density~\cite{marcus}. 
This simple functional form still remains valid at the PM (beyond
mean-field) level for WS-cells of various
geometries~\cite{wennerstrom}, 
although the mean-field prediction for the 
equilibrium boundary density $\bar{n}(R)$ will (in general) 
disagree with the corresponding rigorous PM result due to the 
neglect of intracell microion-microion correlations and 
finite ionic sizes.
Henceforth, to simplify the notation, we will omit the bar to 
denote equilibrium properties. 

\section{Linearization scheme\label{sec:3}}

\subsection{In the presence of neutralizing counterions only}

Let us define a linearized 
free-energy functional that will lead to DH-like 
equations of state for the salt-free model system 
introduced in the previous section.
We start by expanding 
the PB Helmholtz free-energy functional~(\ref{eqn:pbfunct}) about the 
uniform counterion density~(\ref{eqn:uniformdensity}), i.e., 
up to second order in the difference $n(\rb)-\nc$, 
\bea
\beta{\mathrsfs F}_\mathrm{DH}
&=& Z\left[\ln \left(\nc\zeta^3\right)-1\right]
+\frac1{8\pi\lb}\int\dd^3\rb\,\left[\nabla\psi(\rb)\right]^2
+\frac12\nc\int\dd^3\rb \left[\frac{n(\rb)}{\nc}-1\right]^2.\qquad
\label{eqn:dhfunct}
\eea
The linear contribution is absent from~(\ref{eqn:dhfunct}), because
it automatically vanishes,
\be
\int\dd^3\rb \left[\frac{n(\rb)}{\nc}-1\right] = \frac{Z}{\nc}- V=0 .
\label{eqn:linearvanishes}
\ee
As in the previous section, 
we introduce a Lagrange multiplier $\mu_\mathrm{el}$ 
and define the extended Helmholtz free-energy functional
${\mathit{\tilde\Omega}}_\mathrm{DH}=
{\mathrsfs F}_\mathrm{DH}- 
\mu_\mathrm{el}\int\dd^3\rb \,\rho(\rb)$. 
Functional minimization of ${\mathit{\tilde\Omega}}_\mathrm{DH}$
with respect to $n(\rb)$ leads to the 
linearized equilibrium counterion profile,   
\be
n(\rb)= \nc \left[1+\beta\mu_\mathrm{el}-\psi(\rb)\right]. 
\label{eqn:dhprofile}
\ee
It follows 
from~Eqs.~(\ref{eqn:linearvanishes})~and~(\ref{eqn:dhprofile}) 
that the linearized equilibrium Lagrange multiplier
$\beta\mu_\mathrm{el}$ is given by the average electrostatic 
potential inside the WS cell, 
\be
\beta\mu_\mathrm{el}=
\left\langle \psi(\rb) \right\rangle . \label{eqn:dhmuel}
\ee
Inserting the equilibrium counterion-density 
profile~(\ref{eqn:dhprofile}) 
into the Poisson equation~(\ref{eqn:poisson}), leads to the
DH-like equation,
\be
\nabla^2\psi(\rb)=\kappa^2 \left[\psi(\rb)-\beta\mu_\mathrm{el}-1
\right] + \frac{Z\lb}{a^2}\delta^3(|\rb|-a), 
\label{eqn:dhlike}
\ee
where the Debye screening length $\kappa^{-1}$
is evaluated with the 
\textit{uniform counterion density} $\nc$,
\be
\kappa=\kc\equiv\sqrt{4\pi\lb \nc}. \label{eqn:kappa}
\ee
This is different from the usual linearized-PB treatment of the 
spherical WS cell~\cite{alexanderpincus},  
where the Debye screening length is defined in terms 
of the WS-cell boundary density $n(r=R)$.
It should be remarked that the DH-like 
 Eq.~(\ref{eqn:dhlike}) leads 
to a \textit{gauge-invariant} linearized solution, i.e., 
independent of the choice of 
the zero of the potential, because it depends only on the 
difference $\psi(\rb)-\beta\mu_\mathrm{el}$.
This is not the case for the 
commonly used linearized-PB solution for the spherical WS 
cell \cite{footnote4}, when it is disputable how 
to construct a self-consistent 
associated linearized free energy, from which the linearized 
equations may be obtained by a functional minimization. 
 
The solution of the linearized DH-like 
equation~(\ref{eqn:dhlike}) for \textit{an arbitrary} 
WS-cell boundary potential 
$\psi(R)$ and under 
the appropriate boundary conditions,
\be
\left.\frac{\dd\psi(r)}{\dd r}\right|_{r=a}=\frac{Z\lb}{a^2},
\qquad\mbox{ and }\qquad\left.\frac{\dd\psi(r)}{\dd r}\right|_{r=R}=0, 
\ee
reads 
\bea
{\psi} (r)&=&\psi(R)+\frac{2Z\kappa\lb}{\Delta_2(\kappa R,\kappa a)}
-\frac{Z\lb}{r}\,
\frac{\Delta_1(\kappa R,\kappa r)}{\Delta_2(\kappa R,\kappa a)},
\label{eqn:dhsolution}\\
\beta q\lb {E}(r)&=& 
-{\lb}\,\frac{\dd\psi(r)}{\dd r}= -
\frac{Z\lb^2}{r^2}
\frac{\Delta_2(\kappa R,\kappa r)}{\Delta_2(\kappa R,\kappa a)},\\
\Delta_1(u,v)&=&  
\Delta_+(u)\mathrm{e}^{v} - \Delta_-(u)\mathrm{e}^{-v}, \label{eqn:delta1} \\
\Delta_2(u,v)&=&  \Delta_+(u) \Delta_-(v) -
\Delta_-(u)\Delta_+(v), \label{eqn:delta2} \\
\Delta_\pm(u)&=& (1\pm u)\mathrm{e}^{\mp u},
\eea
where the Lagrange multiplier $\beta\mu_\mathrm{el}$ is given by   
\be
\beta\mu_\mathrm{el}=
\left\langle \psi(\rb) \right\rangle 
= \psi(R)+\frac{2Z\kappa\lb}{\Delta_2(\kappa R,\kappa a)}-1.
\ee

Inserting the DH-like solution for the electrostatic 
potential~(\ref{eqn:dhsolution}) 
into the linearized Helmholtz free-energy
functional~(\ref{eqn:dhfunct}) leads, after some algebra, 
to the linearized Helmholtz free 
energy,
\bea
\beta F_\mathrm{DH} &=& 
\beta {\mathrsfs F}_\mathrm{DH} [n(\rb)] =
Z\left[ \ln \left(\nc\zeta^3\right)-\frac32 \right]
+ \frac{Z^2\lb}{2a}\, 
\frac{\Delta_1(\kappa R,\kappa a)}{\Delta_2(\kappa R,\kappa a)}. 
\label{eqn:dhmin}
\eea
In Appendix~\ref{app:b} it is shown that the 
linearized Helmholtz free energy~(\ref{eqn:dhmin}) 
may also be obtained by a Debye charging 
process~\cite{mcquarrie} of the linearized electrostatic 
energy, confirming thus its thermodynamic self-consistency. 

The linearized osmotic pressure (over pure solvent) 
of the colloidal suspension follows from the negative 
\textit{total derivative} of the 
linearized Helmholtz free energy (without including 
the Lagrange multiplier term) with respect to the WS-cell free 
volume $V$,
\bea
\beta P_\mathrm{DH}=
&-&\!\!\left.\frac{\dd\beta F_\mathrm{DH} }{\dd V}\right|_{Z,T}
= \nc -\frac{Z^2\lb}{2a}\, \frac{\dd}{\dd V} 
\left[\frac{\Delta_1(\kappa R,\kappa a)}
{\Delta_2(\kappa R,\kappa a)}\right]_{Z,T}
=\nc\left\{1+ \frac{Z\kappa\lb}{4\Delta^2_2(\kappa R,\kappa a)}
\times\right. \nonumber \\
&\times&\left.
\left[\frac{\Delta_1(\kappa R,\kappa a)}{\kappa a}
\left[ \Delta_1(\kappa R,\kappa a)-\Delta_2(\kappa R,\kappa a)\right]-
4\kappa a \left(1+\frac23 \kappa^2 a^2-\kappa^2 R^2\right)-\frac43\kappa^3 R^3 
\right]\right\},\qquad \label{eqn:dhpressure} 
\eea
where one should take into account both the explicit $R$ dependence as well 
as the volume dependence of the screening length $\kappa^{-1}$ 
when computing the total derivative. 
The first term of~(\ref{eqn:dhpressure}) 
represents the uniform counterion-density ideal-gas law,
while the next term corresponds to the mean-field electrostatic 
corrections \cite{footnote5} due to \textit{intracell}
polyion-microion correlations. 
In Appendix~\ref{app:c} it is shown that the linearized osmotic 
pressure~(\ref{eqn:dhpressure}) can be also obtained by a \textit{formal} 
differentiation of the linearized Helmholtz free energy and that it also 
corresponds to a quadratic expansion 
of the nonlinear PB osmotic pressure~(\ref{eqn:bounddensity}). 
At the end of the next subsection --- that considers the system in the 
presence of added salt ---  we shall find 
that for sufficiently large surface charges 
the linearized osmotic pressure~(\ref{eqn:dhpressure}) is no 
longer a monotonic function of the WS-cell free volume $V$, which 
would imply
a thermodynamical instability and an associated 
phase separation of the system --- in contrast to the full
nonlinear theory, which does not predict any instability. 

\subsection{In the presence of neutralizing counterions and added 
salt (canonical ensemble)}

Let us now add a symmetric monovalent (1:1) salt 
to the system. At the same level of mean-field approximation, all 
microions are treated as inhomogeneous ideal gases, 
described by their average local number densities $n_\pm (\rb)$. 
We will not distinguish 
between counterions and positive ions derived from the salt
dissociation. Therefore $n_+(\rb)$
accounts both for counterions and 
positive salt ions (cations), 
while $n_-(\rb)$ represents the negative coions (anions).
In terms of these number densities, 
the total charge number density reads
\be
\rho(\rb)=n_+(\rb)-n_-(\rb) 
-\frac{Z}{4\pi a^2}\delta^3\left(|\rb|-a\right).
\ee
The effective average uniform densities 
of positive and negative microions are given by 
\be
\bar{n}_\pm =\left\langle n_\pm(\rb) \right\rangle =
\frac{Q_\pm}{V}, \qquad\qquad Q_+ = Z + \ns V, \qquad\qquad
Q_- = \ns V,\label{eqn:uniform}
\ee
where $\ns$ is the \textit{a priori} known effective
average salt concentration and 
$Q_\pm$ are the fixed total number of positive and negative microions
inside a WS cell. Within the cell-model approximation the salt ions are 
evenly distributed between different cells and the average 
salt concentration $\ns$ is the same for each identical WS cell.  
For later convenience,
it will be useful to introduce the dimensionless parameter
\be
s \equiv \frac{Q_-}{Z} = \frac{\ns}{\nc},
\ee 
which measures the contribution of the salt ions to the 
ionic strength in the suspension,
\be
I\equiv \frac12 \left(\nc+2\ns\right) =\frac12 \left(1+2s\right) \nc.
\ee
For example, for $s=1$ there is one pair 
of salt ions for each positive counterion  in solution. 

We expand the nonlinear PB Helmholtz free-energy functional,
\bea
\beta{\mathrsfs F}[n_\pm(\rb)]&=&
\frac1{8\pi\lb}\int\dd^3\rb\,\left[\nabla\psi(\rb)\right]^2
+ \sum_{i=\pm}\int\dd^3\rb\,n_i(\rb)\left\{\ln
  \left[{n_i(\rb)}\zeta^3_i\right]-1\right\} , \quad
\eea
about the average uniform densities~(\ref{eqn:uniform}), up to second order
in the differences $n_\pm(\rb)-\bar{n}_\pm$, 
\bea
\beta{\mathrsfs F}_\mathrm{DH}
&=& \sum_{i=\pm} Q_i\left[\ln \left(\bar{n}_i\zeta^3_i\right)-1\right]
+\frac1{8\pi\lb}\int\dd^3\rb\,\left[\nabla\psi(\rb)\right]^2
+\frac12\sum_{i=\pm}\bar{n}_i
\int\dd^3\rb \left[\frac{n_i(\rb)}{\bar{n}_i}-1\right]^2,\quad
\label{eqn:fdhsalt}
\eea
where $\zeta_\pm$ are the thermal de Broglie wavelengths of 
 cations (including the positive counterions)
and anions, respectively. 
We introduce a Lagrange multiplier $\mu_\mathrm{el}$, which 
will ensure the overall WS-cell charge neutrality, 
\be
\int\dd^3\rb\,\rho(\rb)=0, \qquad\qquad\mbox{ or }
\qquad\qquad
\int\dd^3\rb\,\left[n_+(\rb)-n_-(\rb)\right] = Z,
\label{eqn:chargeneutralitysalt}
\quad
\ee
and define the extended Helmholtz free-energy functional, 
${\mathit{\tilde\Omega}}_\mathrm{DH}={\mathrsfs F}_\mathrm{DH}- 
\mu_\mathrm{el}\int\dd^3\rb\, \rho(\rb)$. 
Functional minimization of ${\mathit{\tilde\Omega}}_\mathrm{DH}$
with respect to $n_\pm(\rb)$  
leads to the equilibrium density profiles, 
\be
n_\pm(\rb)= \bar{n}_\pm \left[1\pm\beta\mu_\mathrm{el}\mp\psi(\rb)\right].
\label{eqn:eqsalt}
\ee
Using~(\ref{eqn:uniform})~and~(\ref{eqn:eqsalt}), 
we obtain that the Lagrange multiplier is given by the average
electrostatic potential inside the WS cell,  
$\beta\mu_\mathrm{el}=\left\langle \psi(\rb) \right\rangle$,  
as in the salt-free case, Eq.~(\ref{eqn:dhmuel}).
Inserting the equilibrium profiles~(\ref{eqn:eqsalt}) 
into the Poisson equation~(\ref{eqn:poisson}),
leads to the DH-like equation,
\bea
\nabla^2\psi(\rb)&=&
\kappa^2 \left[\psi(\rb)-\beta\mu_\mathrm{el} - \frac{1}{1+2s}
\right] + \frac{Z\lb}{a^2}\delta^3(|\rb|-a),
\eea
where the inverse of the Debye screening length is now given by 
\be
\kappa=\sqrt{8\pi\lb I} = \sqrt{4\pi\lb(1+2s)\nc}
= \kc \sqrt{1+2s}.
\ee
Solving the associated linearized DH-like equation leads again 
to the electrostatic 
potential~(\ref{eqn:dhsolution}), with the Lagrange multiplier
for an arbitrary cell-boundary electrostatic
potential $\psi(R)$ given by
\be
\beta\mu_\mathrm{el}= \left\langle \psi(\rb) \right\rangle 
= \psi(R)+\frac{2Z\kappa\lb}{\Delta_2(\kappa R,\kappa a)}- \frac{1}{1+2s} .
\ee 
At this point we should remark that ---  in 
the infinite-dilution limit $(R\to\infty,\nc\to 0)$, 
but in the presence of excess salt $(s\to\infty$, finite $\ns)$ ---
the linearized electrostatic potential  
$\psi(r)$ reduces to the Yukawa screened electrostatic 
potential, 
\be
\lim_{{}^{R\to\infty}_{s\to\infty}}\left[{\psi}(r)-\psi(R)\right]=
-\frac{Z\lb \mathrm{e}^{-\kappa\left(r-a\right)}}
{\left(1+\kappa a\right)r},\qquad \kappa=\sqrt{8\pi\lb\ns},
\ee
which leads to the repulsive electrostatic component 
of the traditional DLVO interaction 
potential\cite{derjaguinlandau,verwey} between two identical 
spherical particles of radius $a$ whose centers are a distance $r$ apart,
\be
W(r) = Z^2 \lb \left(\frac{\mathrm{e}^{\kappa a}}
{1+\kappa a}\right)^2 \frac{\mathrm{e}^{-\kappa r}}{r}. 
\label{eqn:dlvo}
\ee
The phase diagram and dynamical properties of a
suspension of spherical particles interacting 
through the effective 
DLVO pairwise potential~(\ref{eqn:dlvo})
were systematically investigated in Ref.~[\citen{robbins}] 
using molecular and lattice dynamics techniques.
We should note that the polyion-polyion 
interaction potential within the traditional (symmetric) 
DH framework may be obtained from the \textit{exact} 
(non-spherically symmetric) solution of the
Helmholtz equation $\nabla^2\psi(\rb)=\kappa^2 \psi(\rb)$ 
for two spherical charged  particles  \cite{fisherli}. 
The large-separation $(r\to\infty)$ asymptotics of
this pairwise potential leads directly to 
the DLVO interaction potential~(\ref{eqn:dlvo}).  
One should also keep in mind that the 
exact limiting laws (within the RPM) 
 of the underlying DH theory \cite{mcquarrie} --- 
 associated with the Helmholtz equation --- are only valid 
in the asymptotic limit of 
vanishing ionic strengths $(\kappa\to 0)$. 
Most likely this exactness 
does not apply for the asymmetric case of 
strongly charged colloids.
Alternatively, the DLVO interaction potential~(\ref{eqn:dlvo})
may also be obtained by the large-separation asymptotics
of the microion-averaged polyion-polyion potential of mean-force, 
obtained using the MSA integral equation for the polyion- and 
the microion-microion correlations in the PM \cite{medinanoyola}. 

Inserting the equilibrium density profiles~(\ref{eqn:eqsalt})
into the linearized Helmholtz free-energy
functional\linebreak
${\mathrsfs F}_\mathrm{DH} [n_\pm(\rb)]$, 
Eq.~(\ref{eqn:fdhsalt}), 
leads to the linearized Helmholtz free energy, 
\bea
\beta F_\mathrm{DH}=\beta {\mathrsfs F}_\mathrm{DH} [n_\pm(\rb)] &=&  
(1+s)Z \left\{ \ln \left[(1+s)\nc\zeta^3_+\right]-1 \right\}
+ sZ\left[\ln\left(s\nc\zeta^3_-\right)-1\right] 
\nonumber\\&&
+\frac{Z}2 \left[\frac{Z\lb}{a}\, 
\frac{\Delta_1(\kappa R,\kappa a)}{\Delta_2(\kappa R,\kappa a)} - 
\frac1{1+2s}\right]. \label{eqn:freesalted}
\eea
The two first terms of~(\ref{eqn:freesalted}) 
correspond to the ideal-gas entropy 
of the uniform expansion densities $\bar{n}_\pm$, while the last term 
represents the linearized excess Helmholtz free energy due to 
the polarization of the microionic cloud around the polyion.  
In the infinite-dilution limit and in the presence 
of excess salt $(R\to\infty,\nc\to 0,$ but finite $\ns)$, 
\be 
\lim_{{}^{R\to\infty}_{s\to\infty}}
\frac{Z}{2}\left[\frac{Z\lb}{a}\, 
\frac{\Delta_1(\kappa R,\kappa a)}
{\Delta_2(\kappa R,\kappa a)} - \frac1{1+2s}\right]
= \frac{Z^2\lb}{2a\left(1+\kappa a\right)},\qquad
\kappa=\sqrt{8\pi\lb\ns} , \quad \label{eqn:phizeroexcess}
\ee
it coincides with the polyion-counterion interaction free energy
 (including the polyion self-energy) obtained in an extended 
Debye-H\"uckel-Bjerrum approach for colloidal suspensions --- 
 cf.~Eq.~(2)~of~Ref.~[\citen{liquidstate1}].
The asymptotic \textit{electrostatic excess free 
energy}~(\ref{eqn:phizeroexcess}), 
obtained by linearization of the PB WS-cell model functional, 
accidentally \cite{footnote6} 
coincides with the \textit{electrostatic work} done 
in charging up the surface of the polyion against the
ionic atmosphere in the framework of the traditional DH 
theory  --- 
cf. Ref.~[\citen{hill}], pp.339 --- which is 
obtained by a G\"untelberg charging process. 
The G\"untelberg \cite{guntelberg} 
and the Debye \cite{mcquarrie} charging 
processes differ by the fact that in the 
latter the \textit{whole system} (including
the ionic atmosphere) is simultaneously charged.   

The connection between the infinite-dilution 
$(R\to\infty)$ limit of the 
linearized excess  Helmholtz free 
energy~(\ref{eqn:freesalted}), 
\be 
\lim_{{}^{R\to\infty}_{\mbox{\tiny finite }s}}
\frac{Z}{2}\left[\frac{Z\lb}{a}\, 
\frac{\Delta_1(\kappa R,\kappa a)}{\Delta_2(\kappa R,\kappa a)} -
\frac1{1+2s}\right] =
\frac{Z}{2}\left[\frac{Z\lb}{a(1+\kappa a)} 
-\frac1{1+2s}\right],\quad 
\label{eqn:asymphelmholtz}
\ee
and the state-independent 
volume terms obtained by Roij  \textit{et al.}~\cite{roij}
was first reported by Warren, cf.~Eqs.~(7) and (11) 
of Ref.~[\citen{warren}], followed by 
Denton, cf.~Eq.~(55) of Ref.~[\citen{denton}]. 
Subtracting out from~Eq.~(\ref{eqn:asymphelmholtz})
the polyion self-energy, $Z^2\lb/(2a)$,
yields the negative contributions of the state-independent volume 
terms obtained by  Roij  \textit{et al.} --- cf. 
Eq.~(61) of Ref.~[\citen{roij}], 
\be
\frac{Z \np}{2}\left[\frac{Z\lb}{a(1+\kappa a)} 
-\frac1{1+2s}\right] -\frac{Z^2\lb}{2a}\np= -
\frac{Z^2\lb}{2}\,\frac{\np \kappa}{1+\kappa a}-
\frac{2\pi\lb}{\kappa^2}\left(Z \np\right)^2,
\ee
recalling that 
$\np=\left(4\pi R^3/3\right)^{-1}$ is the polyion
density of the suspension. It has been claimed~\cite{roij} that these
volume-dependent (but state-independent) 
 negative  contributions to the Helmholtz 
free energy are responsible for driving a 
gas-liquid phase separation in dilute 
deionized aqueous colloidal suspensions. 

The linearized \textit{canonical} 
osmotic pressure of the colloidal suspension follows from 
the negative \textit{total derivative}
of the linearized Helmholtz free energy (without including 
the Lagrange multiplier term) with respect to the WS-cell free volume $V$,
but keeping fixed the total amount of salt,
\bea
\beta P_\mathrm{DH}^\mathrm{can}(\phi,s) &=&
-\left.\frac{\dd\beta F_\mathrm{DH} }{\dd V}\right|_{Z,T,s}
= (1+2s)\nc -\frac{Z^2\lb}{2a}\, \frac{\dd}{\dd V} 
\left[\frac{\Delta_1(\kappa R,\kappa a)}
{\Delta_2(\kappa R,\kappa a)}\right]_{Z,T,s}
\nonumber\\
&=& (1+2s)\nc\left\{1
+ \frac{Z\kappa\lb}{4(1+2s)\Delta_2^2(\kappa R,\kappa
  a)}\left[\frac{\Delta_1(\kappa R,\kappa a)}{\kappa a}\,\times
\right.\right.\nonumber \\
&&\times \left.\left.
\left[ \Delta_1(\kappa R,\kappa a)-\Delta_2(\kappa R,\kappa a)\right]-
4\kappa a \left(1+\frac23 \kappa^2 a^2-\kappa^2 R^2\right)-\frac43\kappa^3 R^3 
\right]\right\}.\quad \label{eqn:dhpressurecanonical}
\eea
In Appendix~\ref{app:d} it is shown that the linearized canonical 
osmotic 
pressure~(\ref{eqn:dhpressurecanonical}) 
corresponds to a quadratic expansion of the nonlinear PB canonical 
osmotic pressure~(\ref{eqn:pbpressurecanonical}). 

\begin{figure}[hp]
\begin{center}
\epsfig{file=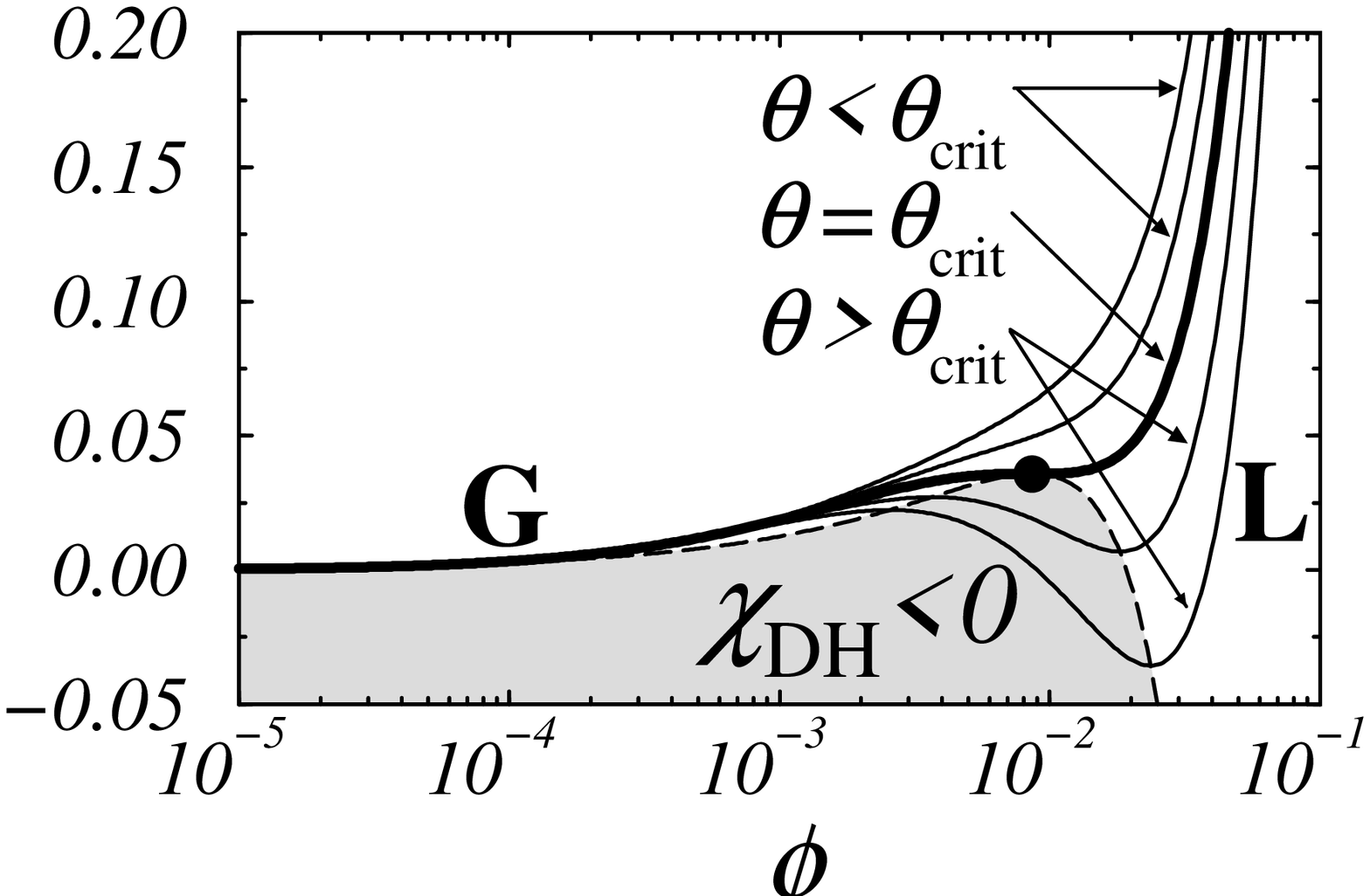,width=0.65\textwidth}
\caption{Salt-free $(s=0)$ linearized osmotic-pressure isotherms
as a function of the volume fraction $\phi=(a/R)^3$. 
From top to bottom the 
isotherms correspond to $\theta=41, 43, 
\theta_\mathrm{crit}=44.902477\cdots \mbox{ (bold line)},47 \mbox{ and } 49$.
In the gray region the salt-free linearized isothermal
compressibility $\chi_\mathrm{DH}$ is negative, 
which  would imply a thermodynamical
instability that leads 
to a phase separation between two fluid phases: 
a low-$\phi$ (dilute) gas (G)
and a high-$\phi$ (dense) liquid (L).
The black circle represents the salt-free critical osmotic pressure
and the dashed curve defines the salt-free spinodal line in 
the $\theta\times\phi$ diagram (the $s=0$ line in 
Figure~\ref{fig:2}).\label{fig:1}}
\epsfig{file=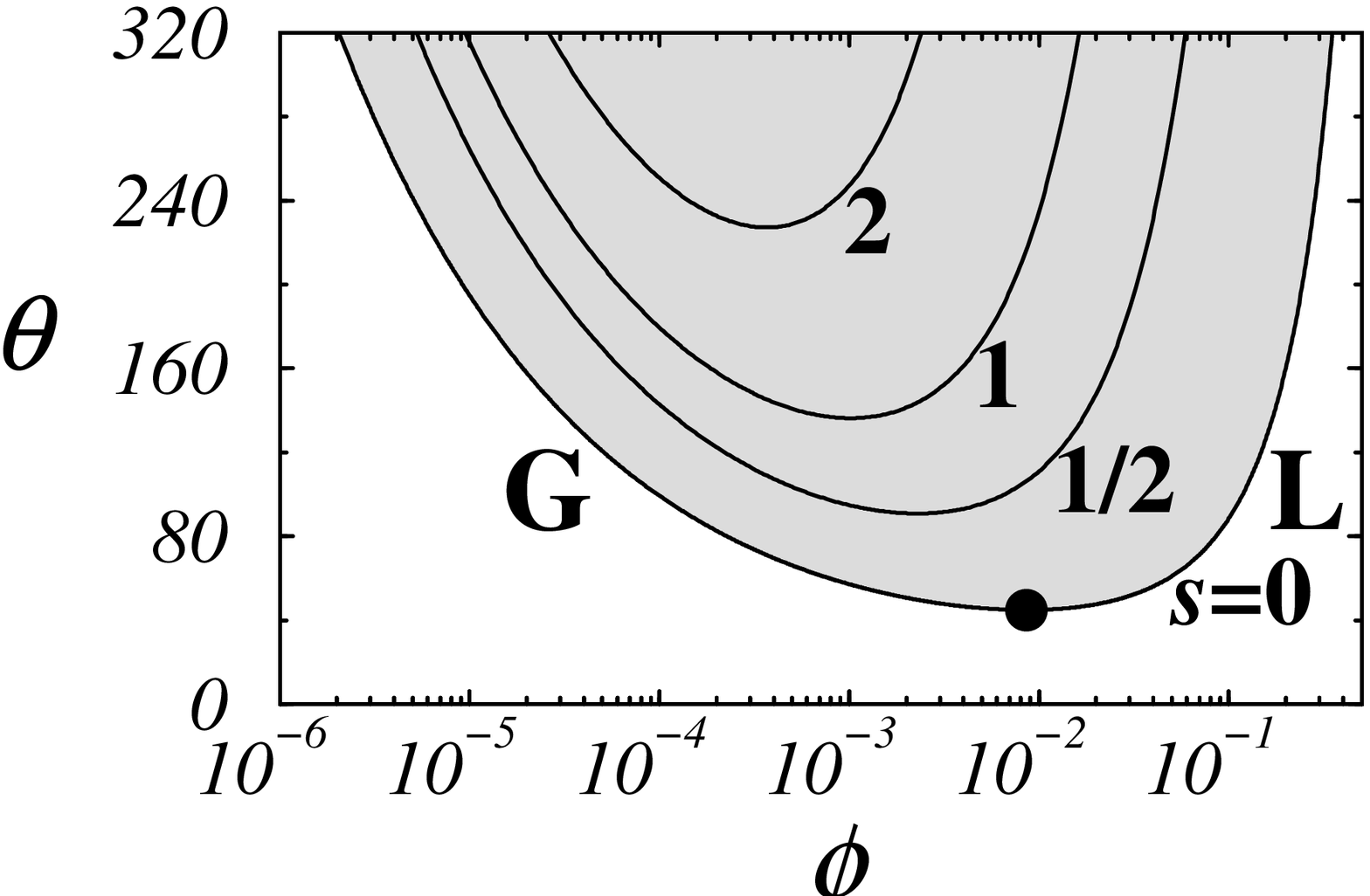,width=0.65\textwidth}
\caption{Spinodal lines associated with the linearized canonical
osmotic pressure $P_\mathrm{DH}^\mathrm{can}$, 
in the $\theta=3Z\lb/a$ versus 
volume fraction $\phi=(a/R)^3$ plane. 
They correspond to lines of vanishing inverse isothermal
compressibility, $\chi_\mathrm{DH}^{-1}=0$.
In the gray region the linearized isothermal
compressibility of the salt-free $(s=0)$ 
suspension becomes negative, leading to 
a coexistence between gas (G) and liquid (L) fluid phases.
Note that this is in contrast to the full nonlinear 
treatment \cite{tellez}, 
which \textit{always} predicts positive compressibilities. 
Addition of monovalent salt reduces the unstable region
by shifting the spinodal lines to higher 
values of $\theta$, as labeled by the different curves 
with increasing values of $s$. 
The black circle represents the salt-free critical 
point (see main text for more details).\label{fig:2}}
\end{center}
\end{figure}

In the vanishing volume fraction of polyions (infinite-dilution) limit, 
$\phi=(a/R)^3\to 0$,
the linearized canonical osmotic pressure has the asymptotic behaviour
\be
\beta P_\mathrm{DH}^\mathrm{can}=
\frac{\theta\left(1+2s\right)\phi}{4\pi a^2 \lb}
\left[1-
\frac{\theta}{10\left(1+2s\right)}\phi^{1/3}-\frac{4\theta^2}{175}\phi^{2/3}
+{\mathrsfs O}\left(\phi\right)\right], \qquad \theta\equiv 
\frac{3Z\lb}{a}, 
\ee
which leads to the asymptotic linearized inverse isothermal 
compressibility,
\be
\beta \chi_\mathrm{DH}^{-1} \equiv 
\np \frac{\dd \beta P_\mathrm{DH}^\mathrm{can}}{\dd \np} 
= Z \np\left(1+2s\right)\left[1-
\frac{2\theta}{15\left(1+2s\right)}\phi^{1/3}-\frac{4\theta^2}{105}\phi^{2/3}
+{\mathrsfs O}\left(\phi\right)\right].
\ee
In contrast to the semi-grand-canonical
case (to be treated in the next subsection), 
the linearized isothermal compressibility in the canonical ensemble 
is stable in the infinite-dilution limit,  
$\lim_{\phi\to 0}  \chi_\mathrm{DH}> 0 $. 
However, as shown in Figures~\ref{fig:1}~and~\ref{fig:2}, 
for finite densities $(\phi\neq 0)$ and 
sufficiently large values of $\theta$,
the linearized canonical 
osmotic pressure $\beta P_\mathrm{DH}^\mathrm{can}$
is no longer a
convex function of the volume fraction $\phi$, 
implying thus the onset of a thermodynamical instability.  
For salt-free suspensions $(s=0)$, 
the associated critical point --- represented by the 
black circle in Figure~\ref{fig:2} --- 
is located at 
\be
\phi_\mathrm{crit}=0.008586189\cdots, 
\qquad
\theta_\mathrm{crit}=44.902477094\cdots,
\label{eqn:critical}
\ee
which is determined by the criticality condition 
$\dd P_\mathrm{DH}^\mathrm{can}/\dd \phi=
\dd^2 P_\mathrm{DH}^\mathrm{can}/\dd \phi^2=0$.
We should stress that the coexistence 
regions between the dilute gas (G)
and the dense liquid (L) phases --- not shown in Figure~\ref{fig:2} ---  
must be determined under the constraints of constant chemical 
potential of polyions $\mup$ and of salt particles $\mus$. 
For example, the salt-free bimodal line that defines the boundary of 
the coexistence region is given by the coupled system of
equations,
\bea
P_\mathrm{DH}^\mathrm{can}(\phi_\mathrm{G},s=0) &=&
P_\mathrm{DH}^\mathrm{can}(\phi_\mathrm{L},s=0), 
\qquad\qquad
\mup(\phi_\mathrm{G},s=0)=\mup(\phi_\mathrm{L},s=0),
\label{eqn:bimodal}
\eea 
where $\mup(\phi,s=0)$ is the salt-free chemical potential 
of the polyions, 
\be
\mup(\phi,s=0) 
\equiv \frac{\dd}{\dd \np}
\left[\frac{F_\mathrm{DH}}{\tilde{V}}\right]_{s=0}
= F_\mathrm{DH}(\phi,s=0) + \frac{P_\mathrm{DH}^\mathrm{can}(\phi,s=0)}
{\np(\phi)}, 
\ee
and $\tilde{V}\equiv {4\pi}a^3/(3\phi) = 1/\np$
is the total WS-cell volume. 
Because the critical behaviour is a spurious result 
of the linearization, it is not worthwhile to 
construct the phase diagrams in detail and we 
restrict ourselves only to present the spinodal lines 
in Figure~\ref{fig:2}. 
This, however, will not change the location of the 
critical points, where the bimodal and the spinodal lines meet. 
Addition of salt \textit{stabilizes} the suspension by 
reducing the unstable region to higher 
values of $\theta$, as also illustrated by the $s\neq 0$ spinodal lines
in Figure~\ref{fig:2}. 
The corresponding bimodal lines are determined by 
generalizing Eqs.~(\ref{eqn:bimodal}) to the case of
 added salt $(s\neq 0)$,
\bea
P_\mathrm{DH}^\mathrm{can}(\phi_\mathrm{G},s) &=&
P_\mathrm{DH}^\mathrm{can}(\phi_\mathrm{L},s), 
\qquad\qquad
\mup(\phi_\mathrm{G},s)=
\mup(\phi_\mathrm{L},s),
\eea 
where the chemical potential of the
polyions in presence of added salt, 
$\mup(\phi,s\neq 0)$,
is given by Eq.~(\ref{eqn:mumacroion}).
With added salt, there is (in the coexistence region) a redistribution of 
microion pairs between the dilute and the dense phases, which still obeys 
the WS-cell charge-neutrality constraint~(\ref{eqn:chargeneutralitysalt}). 
Hence, inside the coexistence region 
 $(\phi_\mathrm{G}\leq\phi\leq\phi_\mathrm{L})$
the two fluid phases will not have the same value 
of $s$ (corresponding to 
the homogeneous system). 
Their values in the gas $(s_\mathrm{G})$ and the liquid $(s_\mathrm{L})$
phases are obtained by imposing the
total conservation of salt particles,
\be
s= \frac{\phi_\mathrm{L}\left(\phi-\phi_\mathrm{G}\right)}
{\phi\left(\phi_\mathrm{L}-\phi_\mathrm{G}\right)} s_\mathrm{L} +
\frac{\phi_\mathrm{G}\left(\phi_\mathrm{L}-\phi\right)}
{\phi\left(\phi_\mathrm{L}-\phi_\mathrm{G}\right)} s_\mathrm{G},
\qquad\qquad \phi_\mathrm{G}\leq\phi\leq\phi_\mathrm{L},
\ee
and the equality in both fluid phases of the chemical potential 
of salt particles $\mus(\phi,s)$ --- as given by~Eq.~(\ref{eqn:musalt}),  
\be
\mus(\phi_\mathrm{G},s_\mathrm{G})=
\mus(\phi_\mathrm{L},s_\mathrm{L}).
\ee

An important and useful concept in charged colloidal 
systems is the charge renormalization \cite{belloni,trizac} of
the polyion bare charge $Z$ for finite volume fractions $\phi$.
For highly charged polyions, i.e. in the $\sigma\equiv Z/(4\pi a^2)
\to\infty$ limit, it has been shown by 
Alexander \textit{et al.} \cite{alexanderpincus} that the 
renormalized effective charge $Z_\mathrm{eff}$ in the 
salt-free system approaches a saturation
value $Z_\mathrm{sat}\approx a w(\phi)/\lb$, 
with $w(\phi)$ assuming numerical 
values  \cite{alexanderpincus,stevens} around 9 to 15
in the volume-fraction range $0.01\leq\phi\leq 0.1$. 
A self-consistent linearized osmotic pressure including charge 
renormalization effects  would  
require the inclusion of additional terms due to the volume-fraction
dependence of the effective charge $Z_\mathrm{eff}=Z_\mathrm{eff}(V)$, 
since the osmotic pressure 
is defined as the negative 
\textit{total derivative} of the Helmholtz free energy with 
respect to the volume $V$. In other words, the linearized canonical 
osmotic pressure $P_\mathrm{DH}^\mathrm{can}$ 
taking into account charge-renormalization
effects \textit{is not simply} given by replacing $Z\to
Z_\mathrm{eff}$ into Eq.~(\ref{eqn:dhpressurecanonical}).
This point will be considered  in a future 
work \cite{rayleigh}.
It is interesting to note, however, that the linearized critical 
threshold $\theta_\mathrm{crit}$ given by Eq.~(\ref{eqn:critical})
is very close to the 
(largest) salt-free saturation $(Z\to\infty)$ effective charge 
$\theta_\mathrm{sat}=3Z_\mathrm{sat}\lb/a\approx 45$ determined by 
Alexander \textit{et al.} \cite{alexanderpincus}.
It has been speculated \cite{yancorrelations}
that this curious coincidence drives suspensions of
highly charged colloids close to criticality. 
This might account for some of the experimental 
findings in dilute deionized aqueous suspensions of highly charged 
colloids \cite{ise,ito,tata}, which then would be explained by  
the presence of strong density fluctuations near the 
criticality. \cite{yancorrelations} 
Another consequence that effective charges are below the saturation
value, and therefore also below the linearized critical threshold, 
is that charge renormalization 
would stabilize the suspension against
phase separation, because the unstable region predicted by 
linearized theory is unreachable (or at least drastically reduced)   
when including renormalized effective charges.
This fact was pointed out 
previously \cite{diehl} using a generalized Debye-H\"uckel-Bjerrum 
approach for charged colloidal suspensions. 
In the present calculation, however, the critical behaviour is 
an artifact of the linearization, which is absent in the 
full nonlinear treatment. 
Therefore, the occurrence of a thermodynamical 
instability can only be explained beyond the WS-cell mean-field 
approximation, by including excluded-volume effects, 
\textit{intercell} polyion-microion and 
microion-microion correlations 
that are neglected in the WS-cell mean-field PB picture. 

\subsection{In contact with an infinite salt reservoir
(semi-grand-canonical ensemble)}

Let us now consider the colloidal suspension in electrochemical
equilibrium with an infinite salt reservoir of fixed bulk density $\nb$. 
The suspension is separated from the infinite reservoir by a
semi-permeable membrane. The solvent and microions 
(counterions and salt ions) can pass through the membrane, 
but not the large polyions. This gives rise
to an imbalance in the osmotic pressure across the semi-permeable
membrane. This equilibrium between the suspension and 
the salt reservoir is referred to as a Donnan 
equilibrium~\cite{donnan,overbeekdonnan,hill,hill1}.
Like in the previous subsections 
we will consider only the case of monovalent counterions 
and symmetric monovalent (1:1) salt. 

The effective average salt concentration in the colloidal suspension, 
$\ns\equiv\left\langle n_-(\rb)\right\rangle$, does not 
coincide with the reservoir bulk density $\nb$ and is not 
known \textit{a priori}. 
A nontrivial question is its dependence with the physical 
parameters of the system, e.g., bulk salt concentration $\nb$,
polyion radius $a$, polyion valence $Z$ and volume fraction 
$\phi\equiv (a/R)^3$.
At the WS-cell PB mean-field level of approximation this problem 
has already been considered in the literature~\cite{reus,tamashiro}
and it is summarized in Appendix~\ref{app:e}.
Compared to the canonical case treated in the 
previous subsection --- when the amount of microions is fixed and 
known \textit{a priori} --- there are two main differences. 
First, because the Donnan equilibrium is established under 
constant electrochemical potential of microions, 
the natural thermodynamical ensemble to perform the calculations is  
the semi-grand-canonical one,
\be
{\mathit{\Omega}}_\mathrm{DH}\equiv {\mathrsfs F}_\mathrm{DH} - 
\sum_{i=\pm}\mu_i\int\dd^3\rb\,n_i(\rb),
\qquad\qquad \beta\mu_\pm=\ln \left(\nb \zeta^3_\pm\right),\quad 
\ee
where we impose the equality of  
the microion electrochemical potentials inside the
colloidal suspension, $\mu_\pm$, to the (mean-field) chemical
potential of ideal gases of uniform density $\nb$ 
for both types of ions in the infinite salt reservoir,
$\beta^{-1}\ln \left(\nb \zeta^3_\pm\right)$.
The second difference is that 
the effective average uniform densities of positive 
and negative ions, about which the linearization should be 
performed, vary in a nontrivial way as the WS-cell free volume $V$
is changed. In Appendix~\ref{app:f} it is shown
that the self-consistent linearized average densities for the 
Donnan problem are 
given by the state-independent zero-th order Donnan densities,
\be
\bar{n}_\pm = 
\frac{\sqrt{\nc^2+(2\nb)^2}\pm \nc}{2}, \label{eqn:donnandensities}
\ee
where $\nc$ is the effective averaged counterion density, defined 
in~(\ref{eqn:uniformdensity}).
These correspond to the uniform densities 
that the system would have in the infinite-temperature
$(\lb=0)$ limit under the constraint of overall 
WS-cell charge neutrality~(\ref{eqn:chargeneutralitysalt}). 
We should remark that they \textit{do not} correspond to the effective
averages of the full nonlinear PB densities~(\ref{eqn:densitydonnan}), 
\be
\left\langle n_\pm(\rb) \right\rangle =
\frac{\sqrt{\nc^2 +
(2\nb)^2\left\langle 
\mathrm{e}^{\psi(\rbs)}\right\rangle\left\langle 
\mathrm{e}^{-\psi(\rbs)}\right\rangle}\pm \nc}2 
= \frac{\sqrt{\nc^2 +
(2\nb)^2 \mathrm{e}^{\left\langle\delta_2(\rbs)\right\rangle} 
+{\mathrsfs O}\left[\left\langle\delta_3(\rb)\right\rangle\right]}\pm
\nc}2 , \label{eqn:densitydonnanaverage}
\ee
because of the nonvanishing quadratic and higher-order $(\nu\geq 2)$ 
contributions of the electrostatic potential differences,
\be
\delta_\nu(\rb)\equiv 
\left[\left\langle \psi\right\rangle-\psi(\rb)\right]^\nu. 
\label{eqn:deltan}
\ee
Here one can see an advantage of the gauge-invariant 
formulation of the problem: the linearized expansion 
densities~(\ref{eqn:donnandensities}) are simply obtained by 
taking the potential-independent, infinite-temperature $(\lb=0)$  
limit of the PB nonlinear averaged 
densities~(\ref{eqn:densitydonnanaverage}), when
the electrostatic potential differences vanish, 
$\lim_{\lb\to 0}\delta_\nu(\rb)\to 0$.
It is also clear that any other choice for the 
expansion densities will not lead to the exact 
potential-independent (infinite-temperature) limit of 
the nonlinear equations. 
As also derived in Appendix~\ref{app:f}, 
the Donnan average densities,
Eqs.~(\ref{eqn:donnandensities}),
do indeed correspond to the functional minimization of the corresponding 
linearized semi-grand-canonical functional. 
We note that Deserno and von Gr\"unberg~\cite{deserno} 
define an optimal linearization point $\bar\psi_\mathrm{opt}$ --- 
related to the above gauge-invariant expansion
 densities by $\bar{n}_\pm=\nb\mathrm{e}^{\mp\bar\psi_\mathrm{opt}}$ ---  
by using arguments based on the plausibility of this choice.
They show that any other choice for the linearization point 
would lead to conflicting inequalities involving 
nonlinear and linearized averages. Finally, we should mention that
even though the uniform expansion densities~(\ref{eqn:donnandensities})
are \textit{internally} 
self-consistent (within the semi-grand-canonical ensemble)  with the 
linearized truncation of the quadratic expansion of the 
nonlinear  semi-grand-canonical functional, 
\textit{global} self-consistency 
(between the canonical and the semi-grand-canonical ensembles)  will 
require the use of the quadratic 
truncation of the nonlinear averages~(\ref{eqn:densitydonnanaverage})
as the self-consistent averaged densities, 
as discussed in detail in Appendix~\ref{app:g}. 
The inclusion of the quadratic contributions into the 
expansion averages, however, do not improve 
the agreement between the linearized and nonlinear equations, 
as can be shown by the explicit 
analytical comparison in the exactly solvable planar case
\cite{plane}. 

Once again, we introduce a Lagrange multiplier $\mu_\mathrm{el}$
that enforces the overall WS-cell  
charge-neutrality~(\ref{eqn:chargeneutralitysalt}),
and define the extended semi-grand-canonical functional, 
${\mathit{\tilde\Omega}}_\mathrm{DH}={\mathit{\Omega}}_\mathrm{DH}- 
\mu_\mathrm{el}\int\dd^3\rb\, \rho(\rb)$. 
Functional minimization of ${\mathit{\tilde\Omega}}_\mathrm{DH}$
with respect to $n_\pm(\rb)$, performed in Appendix~\ref{app:f}, 
leads to the self-consistent linearized averaged 
densities~(\ref{eqn:donnandensities})~and the 
linearized equilibrium profiles, 
\be
n_\pm(\rb)=\bar{n}_\pm \left[1\pm \left\langle\psi(\rb)\right\rangle
\mp \psi(\rb)\right]. \label{eqn:npmdonnan}
\ee
We should remark that, contrary to the salt-free 
and canonical cases, cf.~Eq.~(\ref{eqn:dhmuel}), here 
the equilibrium Lagrange multiplier $\beta\mu_\mathrm{el}$ 
\textit{is not} related to
the average electrostatic potential, 
$\beta\mu_\mathrm{el}\neq \left\langle\psi(\rb)\right\rangle$.

Inserting the linearized equilibrium profiles~(\ref{eqn:npmdonnan})
into the Poisson equation~(\ref{eqn:poisson}),
leads to the DH-like equation,
\bea
\nabla^2\psi(\rb)=\kappa^2 \left[\psi(\rb)-
\left\langle \psi(\rb)\right\rangle -\eta \right] 
+\frac{Z\lb}{a^2}\delta^3(|\rb|-a), 
\label{eqn:dhdonnan}
\eea
where the parameter 
\be
\eta\equiv\frac{\bar n_+ -\bar n_-}
{\bar n_+ +\bar n_-}=\frac{\nc}{\sqrt{\nc^2+(2\nb)^2}},  
\ee
measures the relative importance of the counterions to the ionic
strength in the colloidal suspension,
\be
I\equiv \frac12 \left(\bar{n}_++\bar{n}_-\right) = 
\frac12{\sqrt{\nc^2+(2\nb)^2}} =
\frac{\nc}{2\eta}=\frac{\nb}{\sqrt{1-\eta^2}}.
\ee
Furthermore, the (effective) Debye screening length 
$\kappa^{-1}$ in the colloidal suspension,
\be
\kappa^2 =8\pi\lb I= \frac{\kc^2}{\eta}=\frac{\kb^2}{\sqrt{1-\eta^2}}
> \kb^2,
\ee
is always shorter than the Debye screening 
length $\kb^{-1}\equiv 1/\sqrt{8\pi\lb\nb}$ associated with 
the bulk salt concentration $\nb$ in the reservoir, 
showing that screening is enhanced in the colloidal 
suspension compared to the salt reservoir. 

Solving the associated linearized DH-like 
equation~(\ref{eqn:dhdonnan}) leads again 
to the electrostatic 
potential~(\ref{eqn:dhsolution}), 
with the average electrostatic potential 
over the WS-cell volume for an arbitrary cell-boundary electrostatic
potential $\psi(R)$ given by
\be
\left\langle \psi(\rb) \right\rangle 
=\psi(R)+ \frac{2Z\kappa\lb}{\Delta_2(\kappa R,\kappa a)}- \eta.
\ee

In close analogy to the linearized Helmholtz free energy for 
the canonical case, Eq.~(\ref{eqn:freesalted}), 
the final expression for the linearized semi-grand-canonical potential is 
\bea
\beta \Omega_\mathrm{DH}=\beta{\mathit\Omega}_\mathrm{DH}
[n_\pm(\rb)]&=&
\sum_{i=\pm} V \bar{n}_i\left[\ln \left(\frac{\bar{n}_i}{\nb}\right)-1\right]
+ \frac{Z}{2}\left[\frac{Z\lb}{a}\, 
\frac{\Delta_1(\kappa R,\kappa a)}{\Delta_2(\kappa R,\kappa a)} -
\eta\right] \nonumber\\&=& 
Z \left[\arctanh\eta-\frac1{\eta}-\frac{\eta}2
+ \frac{Z\lb}{2a}\, 
\frac{\Delta_1(\kappa R,\kappa a)}
{\Delta_2(\kappa R,\kappa a)}\right]. 
\qquad \label{eqn:omegadh}
\eea

The  linearized  \textit{semi-grand-canonical}
osmotic pressure follows from the negative 
\textit{total derivative} of the 
linearized semi-grand-canonical potential 
 (without including the Lagrange multiplier term
associated with $\mu_\mathrm{el}$) with respect to the WS-cell free 
volume $V$, but keeping fixed the microion chemical potentials  
 $\mu_\pm=\beta^{-1}\ln\left(\nb\zeta^3_\pm\right)$,
\bea
\beta P_\mathrm{DH}^\mathrm{sgc}(\phi,\nb)
&=&\left.-\frac{\dd\beta\Omega_\mathrm{DH}}
{\dd V}\right|_{Z,T,\mu_\pm}
%
=\frac{\nc}{\eta}
\left\{1+\frac{\eta^2}2(\eta^2-1)
+ \frac{Z\kappa\lb\eta^3}{4\Delta_2^2(\kappa R,\kappa a)}
\left[\frac{\Delta_1(\kappa R,\kappa a)}{\kappa a}\times
\right.\right. \nonumber \\
&&\times
\left.\left.
\left[ \Delta_1(\kappa R,\kappa a)-\Delta_2(\kappa R,\kappa a)\right]-
4\kappa a \left(1+\frac{2\kappa^2a^2}
{3\eta^2}-\kappa^2 R^2\right)-4\left(1-\frac2{3\eta^2}\right)\kappa^3 R^3 
\right]\right\}, \qquad\label{eqn:donnandhpressure}
\eea
where the prefactor of the right-hand side represents the 
ideal-gas Donnan osmotic pressure taking only the WS-cell charge 
neutrality~(\ref{eqn:chargeneutralitysalt}) into account,
\be
\frac{\nc}{\eta}= \bar{n}_+ + \bar{n}_- =\sqrt{\nc^2+(2\nb)^2}. 
\ee
In analogy to the salt-free, Eq.~(\ref{eqn:dhpressure}), 
and the canonical, Eq.~(\ref{eqn:dhpressurecanonical}),
cases, the first term 
of~(\ref{eqn:donnandhpressure}) represents the ideal-gas law
associated to the state-independent 
zero-th order Donnan densities, while the remaining terms correspond to the 
mean-field electrostatic corrections due to the 
microionic polarization around the polyion. 
We should note that the second term inside curly brackets 
depends only on $\eta$ and is thus $\lb$ independent. 
This could suggest that the  zero-th order Donnan ideal-gas 
law, $\lim_{\lb\to 0}
\beta P_\mathrm{DH}^\mathrm{sgc}=\nc/\eta$, would not be recovered 
in the weak-coupling limit.
This $\lb$-independent term, however, is indeed necessary 
to cancel the contributions    
that arise from the last term in the $\lb\to 0$ limit in order 
to give the correct potential-independent infinite-temperature limit.  
Moreover, it is shown in Appendix~\ref{app:e} that the 
linearized semi-grand-canonical  
osmotic pressure~(\ref{eqn:donnandhpressure}) corresponds
to a quadratic expansion of the nonlinear 
semi-grand-canonical  osmotic pressure~(\ref{eqn:donnanpbpressure}). 

The linearized osmotic-pressure difference between 
the colloidal suspension and the infinite salt reservoir obeys 
$\beta\Delta P_\mathrm{DH}\equiv \beta P_\mathrm{DH}^\mathrm{sgc}-2\nb=
\beta P_\mathrm{DH}^\mathrm{sgc}-({\nc}/{\eta})\sqrt{1-\eta^2}$,
with $\beta P_\mathrm{DH}^\mathrm{sgc}$ 
given by~Eq.~(\ref{eqn:donnandhpressure}).
In the next section we show that the  
 linearized semi-grand-canonical
osmotic-pressure difference $\beta\Delta P_\mathrm{DH}$ 
is intrinsically thermodynamically unstable in 
the infinite-dilution limit and we compare 
with expressions previously obtained by  
Deserno and von Gr\"unberg~\cite{deserno}. 


\section{Comparison with Deserno and von Gr\"unberg results\label{sec:4}}

As already pointed out previously in the 
literature \cite{grunberg,deserno}, 
in this section we will show that the 
linearized semi-grand-canonical osmotic pressure 
$P_\mathrm{DH}^\mathrm{sgc}$ 
defined by Eq.~(\ref{eqn:donnandhpressure})
yields artifacts in the low-temperature, 
high-surface charge or infinite-dilution (of polyions)
limits.
In contradiction to the exact nonlinear 
result~(\ref{eqn:donnanpbpressure}), which yields 
an osmotic-pressure difference that is always 
positive \cite{footnote7}, 
 $\beta\Delta P=\beta P-2\nb>0$, the linearized version
$\Delta P_\mathrm{DH}$ becomes negative in the above mentioned limits.
In an attempt to define the osmotic pressure in a linearized 
framework, Deserno and von Gr\"unberg \cite{deserno} introduced an 
alternative definition, $P_1$,
that has the advantage of being exempt from any instabilities 
and is obtained via the \textit{partial derivative} 
of the semi-grand-canonical potential with respect to the volume,
keeping the optimal linearization point $\bar\psi_\mathrm{opt}$
(to be defined below) fixed. 
Their second osmotic pressure definition
coincides with the linearized version~(\ref{eqn:donnandhpressure}) 
obtained in the previous section, $P_2\equiv P_\mathrm{DH}^\mathrm{sgc}$,
and is obtained via the \textit{total derivative} 
of the semi-grand-canonical potential with respect to the volume.
These two distinct osmotic-pressure definitions are given, respectively, 
by Eqs.~(43)~and~(44) of Ref.~[\citen{deserno}] for $d=3$, 
\bea
\frac{\beta P_1}{2\nb}&=& 
1+\frac{\left(\cosh\bar\psi_\mathrm{opt}-1\right)^2}
{2\cosh\bar\psi_\mathrm{opt}} + 
\frac{\sinh^2\bar\psi_\mathrm{opt}}{2\cosh\bar\psi_\mathrm{opt}}
\left(\frac{1-\phi}{3{\cal D}\sqrt{\phi}}\right)^2 \geq 1, \\
\frac{\beta P_2}{2\nb}&=&  \frac{\beta P_1}{2\nb} - 
\frac{\sinh^4\bar\psi_\mathrm{opt}}{2\cosh^3\bar\psi_\mathrm{opt}}
\left\{\frac{1-\phi}{6\phi}\left[\frac1{{\cal D}^2}-
\kappa a \frac{\cal E}{\cal D}+
\kappa^2 a^2\left(1-\frac{{\cal E}^2}{{\cal D}^2}
\right)\right]-1\right\}, \qquad\\
{\cal D}&=& I_{3/2}(\kappa R)\,K_{3/2}(\kappa a) -
K_{3/2}(\kappa R)\,I_{3/2}(\kappa a), \\
{\cal E}&=& I_{3/2}(\kappa R)\,K_{1/2}(\kappa a) + 
K_{3/2}(\kappa R)\,I_{1/2}(\kappa a),  
\eea
where $\left\{I_\nu,K_\nu\right\}$ are 
the modified Bessel functions \cite{abramowitz} of the first and the 
second kind, respectively, and the optimal linearization point
$\bar\psi_\mathrm{opt}$ satisfies the relations
\be
\tanh \bar\psi_\mathrm{opt}=-\eta, \qquad
\cosh\bar\psi_\mathrm{opt}=\left(\frac{\kappa}{\kb}\right)^2
=\frac{1}{\sqrt{1-\eta^2}},
\qquad \sinh\bar\psi_\mathrm{opt}=-\frac{\nc}{2\nb}=
-\frac{\eta}{\sqrt{1-\eta^2}}. 
\ee
In accordance with Eqs.~(23)~and~(26) of Ref.~[\citen{deserno}], 
they can be recast in a simpler formal form in terms of the 
gauge-invariant electrostatic potential differences $\delta_\nu(\rb)$,
defined by~Eq.~(\ref{eqn:deltan}), 
\bea
\beta P_1 &=&
\frac{\nc}{\eta}
\left[1-\frac{\eta^2}2+\frac{2Z^2\kappa^2\lb^2}
{\Delta^2_2(\kappa R,\kappa a)} \right]= 
\frac{\nc}{\eta}
\left[1+\eta\delta_1(R)
+\frac12\delta_2(R)\right]
,\qquad \\
\beta P_2&=& 
\frac{\nc}{\eta}
\left[1+\eta\delta_1(R)
+\frac12\delta_2(R)
-\frac{\eta^2}2 \left\langle\delta_2(\rb)
\right\rangle \right],\qquad
\eea
from which one can see that they differ by a 
term that is quadratic in the electrostatic-potential difference.
Looking at Eq.~(\ref{eqn:averagedelta2}) of Appendix~\ref{app:c},
one may trace back that the omitted contribution in $P_1$  
originates from the volume dependence of the optimal linearization
point $\bar\psi_\mathrm{opt}$, 
in accordance to the interpretation given by Deserno and 
von Gr\"unberg \cite{deserno} for the two distinct pressure definitions. 
We should recall that the 
linearized semi-grand-canonical osmotic-pressure~(\ref{eqn:donnandhpressure}) 
coincides with the second pressure definition, 
$P_\mathrm{DH}^\mathrm{sgc}\equiv P_2$, as shown in Appendix~\ref{app:e} by 
a quadratic expansion of the nonlinear osmotic pressure. It
corresponds indeed to the negative \textit{total derivative} of the 
linearized semi-grand-canonical potential $\Omega_\mathrm{DH}$ with 
respect to the WS-cell free volume $V$, which we thus believe to be 
the consistent and correct definition of the osmotic pressure.

It is convenient to introduce the dimensionless
linearized osmotic-pressure differences, 
\be
\Pi_i \equiv \frac{\beta\Delta P_i}{2\nb} = \frac{\beta P_i}{2\nb}-1, 
\qquad i=1,2.
\ee
In the vanishing volume fraction of polyions (infinite-dilution) limit, 
$\phi=(a/R)^3\to 0$, we may write 
the asymptotic linearized osmotic-pressure differences in terms 
of $\theta\equiv 3Z\lb/a$ and $\hat{a}\equiv \kb a$,
\bea
\Pi_1 &=& \frac{\theta^4\phi^4}{8\hat{a}^8}
%
%
+{\mathrsfs O}\left[\phi^5,\theta^2\phi^{2/3}
\exp\left(-2\hat{a}\phi^{-1/3}\right)\right],
\qquad\\
\Pi_2 &=& 
-\frac{\theta^4\phi^3}{12\hat{a}^5(1+\hat{a})^2}
-\frac{5\theta^4\phi^4}{8\hat{a}^8}\left[
\frac{2\hat{a}^3}{5 \left(1 +\hat{a}\right)^2}-1\right]
+{\mathrsfs O}\left[\phi^5,\theta^2
\phi^{2/3}\exp\left(-2\hat{a}\phi^{-1/3}\right)\right],\quad
\label{eqn:pi2asympt}
%
%
\eea
which lead to the asymptotic linearized inverse isothermal 
compressibilities,
\bea
\beta \chi_1^{-1} \equiv 
2\nb 
\np \frac{\dd\Pi_1}{\dd\np} 
&=& \frac{Z\np\theta^3\phi^3}{2\hat{a}^6}+{\mathrsfs
  O}\left[\phi^4,\theta \phi^{-2/3}
\exp\left(-2\hat{a}\phi^{-1/3}\right)\right],\\
\beta \chi_2^{-1} \equiv
2\nb 
\np \frac{\dd\Pi_2}{\dd \np}
&=& -Z \np\theta^3\left\{
\frac{\phi^2}{4\hat{a}^3(1+\hat{a})^2}
+\frac{5\phi^3}{2\hat{a}^6}\left[\frac{2\hat{a}^3}
{5 \left(1+\hat{a}\right)^2}-1\right]\right\}
\qquad\nonumber\\&&
+{\mathrsfs O}\left[\phi^4,\theta \phi^{-2/3}
\exp\left(-2\hat{a}\phi^{-1/3}\right)\right].
\eea
While neglecting the contribution of the last quadratic term 
in the linearized osmotic pressure $P_1$ always leads 
to positive isothermal compressibilities, $\lim_{\phi\to 0}
\chi_1>0$, its inclusion in $P_2$ \textit{always} yields  
negative isothermal compressibilities in the infinite-dilution
limit for nonvanishing $\lb$, $\lim_{\phi\to 0}
\chi_2<0$. 
This means that the pressure definition $P_2$ predicts that 
the infinite-dilution phase is 
unstable, in contrast to the 
canonical case, as shown in Figure~\ref{fig:3}. 
Therefore the \textit{thermodynamically consistent} 
linearized osmotic pressure $P_2$
\textit{intrinsically fails at infinite dilution,} leading 
to negative isothermal compressibilities in this limit. 
This fact was first noticed for the spherical case 
in Ref.~[\citen{grunberg}] and generalized to WS cells in 
$d$-dimensions in Ref.~[\citen{deserno}].  
For finite densities,  in contrast to the canonical 
case (Figure~\ref{fig:2}),
the low-$\phi$ dilute (gas) phase may only be stable for sufficiently
small bulk salt concentrations, as shown in
Figure~\ref{fig:3}.
For sufficiently large $\kb a$ the 
finite-temperature critical point --- located in the vicinity
of the salt-free critical point (the black circle in 
Figure~\ref{fig:3}) --- disappears.  
Note the nonmonotonic behaviour of the spinodal lines 
associated to $P_2$ for $\kb a=10^{-1}$ and 
$\kb a=10^{2}$ (inset), which leads to 
oscillating osmotic pressures. We should note, however, that the 
WS-cell model ceases to be meaningful for such high volume 
fractions in the latter case of strong screening.
Finally we should remark that \textit{beyond} the 
linearized PB WS-cell model approximation the thermodynamical 
instability at the infinite-dilution limit will be removed by 
taking into account the translational entropy of the polyions, 
which yields an osmotic-pressure contribution that is linear in $\phi$
and, therefore, overcomes the negative cubic leading term in 
the asymptotic linearized osmotic-pressure difference~(\ref{eqn:pi2asympt}).
However, because our analysis restricts to the linearization 
of the PB WS-cell model, the effect of this stabilizing entropic 
contribution --- which may drastically alter the spinodal lines,
specially in the low-volume fraction region  --- is not considered here. 
Note that the PB WS-cell model, in its full nonlinear version, is 
 fully stable \cite{tellez} even without 
invoking this stabilizing contribution. 

Let us stress again that thermodynamic consistency 
and stability are independent concepts. This can be 
illustrated by inspecting the two linearized osmotic-pressure 
definitions proposed in Ref.~[\citen{deserno}]. 
The linearized osmotic pressure $\Pi_2$, although 
not fully stable, is self-consistent with quadratic 
expansions of the nonlinear osmotic pressure.
The unstable region of $\Pi_2$ just reflects the 
breakdown of the linearization scheme to the nonlinear
PB equation. On the other hand, $\Pi_1$, that displays positive
isothermal compressibilities in the infinite-dilution limit
and does not present any instabilities, 
is not \textit{thermodynamically consistent,} 
its stability being purely fortuitous. 
This can be further fortified by 
comparing the \textit{exact} analytical expressions of 
the nonlinear osmotic pressure for the planar geometry \cite{plane} 
with the two corresponding linearized versions. 
In the subsequent paper we show that both linearized expressions, $\Pi_1$
and $\Pi_2$, approach asymptotically the exact nonlinear result 
in the appropriate (weak-coupling, $\lb\to 0$) limit. Their
convergence, however, are very different and it is the self-consistent
definition $\Pi_2$ that gives a better approximation to the full
nonlinear equation. Although analytical proofs can only be obtained 
for the planar case, we believe the same argument applies for any geometry.

\section{Concluding remarks\label{sec:5}}

We performed a linearization scheme consistent with quadratic
expansions of the appropriate nonlinear thermodynamic functional. 
By using gauge-invariant forms of the electrostatic potential,
we have shown that the linearized osmotic pressures correspond 
to quadratic expansions of the corresponding nonlinear versions
for the three cases investigated: in the presence of neutralizing 
counterions only (salt-free case), in the presence of fixed
amount of added salt (canonical case) and in electrochemical equilibrium 
with an infinite salt-reservoir (Donnan equilibrium, 
semi-grand-canonical case).

Contrary to previous works \cite{grunberg,deserno}, we adopted
a gauge-invariant formulation with the inclusion of a Lagrange 
multiplier term to account for the charge-neutrality constraint. 
In the case of the Donnan equilibrium, it is shown that the 
minimization of the associated linearized semi-grand-canonical
functional leads indeed to the state-independent zero-th order Donnan 
densities as the self-consistent expansion averages for the linearization. 
Therefore the optimality of the optimal expansion point 
 $\bar\psi_\mathrm{opt}$ introduced 
by Deserno and von Gr\"unberg \cite{deserno} can be understood  
as corresponding to a self-consistent minimization of the 
linearized functional.

It is shown that the self-consistent linearized osmotic pressure
in the semi-grand-canonical ensemble --- 
as already pointed out in the literature \cite{grunberg,deserno} --- 
leads to artifacts  
in the infinite-dilution, high-surface charge and strong-coupling 
limits, where it predicts negative osmotic-pressure differences
between the colloidal suspension and the salt reservoir and 
negative isothermal compressibilities.
Attempts to define a fully stable linearized 
equation of state based on the partial 
derivative of the linearized semi-grand-canonical potential with respect to
the volume  \cite{deserno} can not be justified in our approach
based on the minimization of the linearized functional, 
its stability being a fortuitous result. 
This can be seen most clearly in the analytically tractable case 
of two infinite charged planes in electrochemical equilibrium 
with an infinite salt reservoir, which we present in a companion 
paper \cite{plane}.

To avoid confusion we should stress at this point the exactness of
the PB nonlinear solution at the mean-field level, 
its range of validity and limitations. 
In this work we discussed the linearization procedure 
in the framework of the nonlinear PB and the WS-cell model.
The linearization constitutes here an approximation to the 
nonlinear treatment, whose exact results (at mean-field level) 
may be then compared to the linearized ones, allowing a control over 
the approximations and the onset of possible artifacts 
introduced by the linearization. Of course, we are not able 
to predict correct results for real systems when the (starting) 
nonlinear theory itself breaks down. In this case, 
an eventual linearized result may \textit{accidentally} lead to a 
correct prediction of, say, phase separation, 
due to the simultaneous application 
of two inadequate approximations, namely, the mean-field PB 
equation and its subsequent linearization. 
The fact that PB nonlinear theory for WS-cell models  
always leads to stable suspensions does not invalidate
phase separation in real systems, 
which may be due to finite-size effects, \textit{intra- and intercell} 
 microion-microion and \textit{intercell} polyion-microion correlations 
that are neglected in the WS-cell mean-field PB picture.
In the Donnan-equilibrium case, one should also take
the microion-microion correlations in the infinite salt reservoir  
into account. 
However, the phase separation in deionized aqueous 
suspensions of charged colloids  predicted by 
linearized theories 
 \cite{roij1,roij,hansen,warren,denton,chan,petris,bhuiyan}
 --- which are claimed \cite{roij1,roij} to be driven 
by the state-independent volume 
terms \cite{beresford,chanpre}   ---  seems to be related 
to mathematical artifacts of the 
linearization itself  and does not correspond 
thus to a real physical effect, as already  
pointed out in previous works \cite{grunberg,deserno} 
via the linearization of the PB functional 
for WS-cell models and revisited it here. 

\section*{Acknowledgments}

The authors are grateful to M.~Deserno and Y.~Levin for 
thoughtful and intensive discussions, in particular to the former for sharing
his results on the Donnan linearization prior to publication.
Discussions with H.~H.~von Gr\"unberg and E.~Trizac are also acknowledged.    
We would also like to thank the financial support
by the Max-Planck-Gesellschaft
and the Alexander von Humboldt-Stiftung. 

\appendix
\setcounter{equation}{0}
\renewcommand{\thesection}{\Alph{section}}
\renewcommand{\theequation}{\Alph{section}\arabic{equation}}
\section{Nonlinear osmotic-pressure boundary density theorem\label{app:a}}

In this Appendix it will be shown that the salt-free
nonlinear osmotic pressure is related to the counterion
density at the WS-cell boundary --- 
see also Section~3~of Ref.~[\citen{desernohouches}]
for the fixed-gauge derivation.
For the Donnan problem, the fixed-gauge derivation 
is presented in Section II.C of Ref.~[\citen{deserno}].

The osmotic pressure (over the pure solvent) $P$ is defined as 
the negative total derivative of the Helmholtz
{free energy} $F$ --- the functional 
${\mathrsfs F}$ evaluated at the equilibrium profile
${n}(\rb)=\bar{n}(\rb)$ ---  with respect to the WS-cell free volume $V$,
\bea
P &=& -\frac{\dd F}{\dd V}
= -\frac{\dd{\mathrsfs F}[\bar{n}(\rb)]}{\dd V} . 
\eea
The calculation of the osmotic pressure starting from the 
explicit form of the optimized Helmholtz free 
energy $F$, Eq.~(\ref{eqn:pb}), turns out to be 
nontrivial, because both the equilibrium counterion 
profile $\bar{n}(\rb)$ as well as the Lagrange multiplier 
$\beta\bar\mu_\mathrm{el}$ 
depend implicitly on the WS-cell free volume $V$. 
However, because 
$\beta\mu_\mathrm{el}$ was introduced as a Lagrange multiplier to
enforce the charge neutrality, it is much easier to 
consider the total derivative with respect to the volume 
of the extended Helmholtz free-energy \textit{functional} 
${\mathit{\tilde\Omega}}[n(\rb)]= {\mathrsfs F}-\mu_\mathrm{el}\int\dd^3\rb\, 
\rho(\rb)$
 --- but now considering the Lagrange multiplier 
$\mu_\mathrm{el}$ \textit{independent} 
of the WS-cell free volume $V$ --- 
evaluated at the optimized profile ${n}(\rb)=\bar{n}(\rb)$,
\bea
\frac{\dd{\mathrsfs F}[\bar{n}(\rb)]}{\dd V} &=&
\left.\frac{\delta{\mathit {\tilde\Omega}}[{n}(\rb)]}{\delta V}
\right|_{\bar{n}(\rbs)} =
 \left.\frac{\partial {\mathit {\tilde\Omega}}[{n}(\rb)]}{\partial V }
\right|_{\bar{n}(\rbs)} +
\int\dd^3\rb'\left.\frac{\delta
{\mathit {\tilde\Omega}}[{n}(\rb)]}{\delta n(\rb')}\right|_{\bar{n}(\rbs')}
\frac{\dd\bar n(\rb')}{\dd V}, \qquad
\eea
where the derivatives of ${\mathit {\tilde\Omega}}$  
are taken with constant $\mu_\mathrm{el}$.
Because ${\mathit {\tilde\Omega}}[{n}(\rb)]$ is stationary with respect to the 
optimized profile $\bar n(\rb)$, 
\be
\left.\frac{\delta
{\mathit {\tilde\Omega}}[{n}(\rb)]}{\delta n(\rb')}\right|_{\bar{n}(\rbs')} = 0,
\ee
only the partial-derivative term contributes to the osmotic pressure. 
Furthermore, because the only 
explicit dependence of ${\mathit {\tilde\Omega}}[{n}(\rb)]$ comes from the 
integration limit \cite{desernohouches}, we are lead to 
\bea
\beta P &=& 
 -\frac{\dd\beta {\mathrsfs F}[\bar{n}(\rb)]}{\dd V}
= -\left.\frac{\partial {\beta \mathit {\tilde\Omega}}[{n}(\rb)]}{\partial V }
\right|_{\bar{n}(\rbs)}
= -\left.\frac{\partial {\beta {\mathrsfs F}}[{n}(\rb)]}{\partial V }
\right|_{\bar{n}(\rbs)} + \left[
\beta\mu_\mathrm{el}\,\frac{\partial}{\partial V }\!\int\dd^3\rb\, n(\rb)
\right]_{\bar{n}(\rbs)} \nonumber\\
&=& -\bar\psi(R)\bar{n}(R) -\bar{n}(R)
\left\{\ln\left[\bar{n}(R)\zeta^3\right]-1\right\} 
+ \beta\bar\mu_\mathrm{el}\,\bar{n}(R)= \bar{n}(R),
\eea
which is the mean-field result \cite{marcus} that
relates the salt-free nonlinear osmotic pressure
to the counterion density at the WS-cell boundary 
when intracell microion-microion correlations are 
neglected. 

\section{Linearized free energy from a 
Debye charging process\label{app:b}}
\setcounter{equation}{0}

In this Appendix it is shown that the 
linearized Helmholtz free energy can also be obtained by  
a Debye charging process~\cite{mcquarrie} 
of the linearized electrostatic 
energy. This is obtained by 
inserting the DH-like solution~(\ref{eqn:dhsolution}) 
into~(\ref{eqn:elec_energy}),
\bea
\beta U(\kappa,\lb) &=& \frac1{8\pi\lb}\int\dd^3\rb\,
\left[\nabla\psi(r)\right]^2 \nonumber\\  
&=& \frac{Z^2 \lb}{2a\Delta_2^2(\kappa R,\kappa a)}
\left[\Delta_1(\kappa R,\kappa a)\Delta_2(\kappa R,\kappa a)-\kappa^2a
\int_a^R \dd r\, \Delta_1^2(\kappa R,\kappa r) \right]. \label{eqn:linear_u}
\eea
The linearized excess Helmholtz free energy over the ideal-gas 
entropy of the reference uniform state is then 
obtained by a Debye charging process~\cite{mcquarrie}
of the linearized electrostatic energy~(\ref{eqn:linear_u}),
\be
\beta {F}_\mathrm{DH}   - 
Z\left[\ln \left(\nc\zeta^3\right)-1\right]
= 2\int_0^1 \dd \lambda\,\frac{\beta U(\lambda\kappa,
\lambda^2\lb)}{\lambda} =
\frac{Z^2\lb}{2a}\left[\frac{\Delta_1(\kappa R,\kappa a)}
{\Delta_2(\kappa R,\kappa a)}-
\frac{3a}{\kappa^2\left(R^3-a^3\right)}\right], 
\ee
which regains Eq.~(\ref{eqn:dhmin}),
obtained by adding the linearized 
electrostatic energy~(\ref{eqn:linear_u})
to the quadratic expansion of the
ideal-gas entropy.

\section{Formal derivation of the 
linearized osmotic pressure\label{app:c}}
\setcounter{equation}{0}

In this Appendix it will be shown that the salt-free linearized  
osmotic pressure is given by a quadratic expansion 
of the nonlinear counterion density at the WS-cell boundary,
\bea
\beta P_\mathrm{DH}&=&-\frac{\dd\beta F_\mathrm{DH}}{\dd V}=
-\frac{\dd\beta{\mathrsfs F}_\mathrm{DH}[\bar{n}(\rb)]}{\dd V}
\nonumber\\&=& 
 \nc\left\{1+\left\langle\bar\psi\right\rangle-\bar\psi(R)+\frac12
\left[\left\langle\bar\psi\right\rangle-\bar\psi(R)\right]^2
-\frac12\left\langle\vphantom{\frac12}
\left[\left\langle\bar\psi\right\rangle-
\bar\psi\right]^2\right\rangle\right\}, 
\qquad \label{eqn:dhpressuredirect}
\eea
which may be obtained by truncating the expansion of 
the salt-free nonlinear PB osmotic pressure up to 
the quadratic terms, 
\bea
\beta P &=& 
\frac{\nc\exp\left[{\left\langle\bar\psi\right\rangle-\bar\psi(R)}\right]}
{\left\langle\exp\left[{\left\langle\bar\psi\right\rangle-\bar\psi(\rb)}
\right]\right\rangle} = \beta P_\mathrm{DH} 
+{\mathrsfs O}\left(
\left[\left\langle\bar\psi\right\rangle-\bar\psi(R)\right]^3,
\left\langle\vphantom{\frac12}
\left[\left\langle\bar\psi\right\rangle-\bar\psi\right]^3\right\rangle
\right),\qquad \label{eqn:pbdhpressure}
\eea
where the bar denotes equilibrium properties. 
One route to check that the salt-free linearized 
osmotic pressure~(\ref{eqn:dhpressure}) may be cast in the 
form~(\ref{eqn:dhpressuredirect}) is by using 
the linearized DH-like solution to the 
electrostatic potential~(\ref{eqn:dhsolution}) and computing 
explicitly the 
spatial averages 
\bea
\left\langle \frac{\Delta_1(\kappa R,\kappa r)}{r}\right\rangle&=&
\frac{4\pi}{V}\int_a^R\dd r\, r \,\Delta_1(\kappa R,\kappa r) = 
-\frac{4\pi}{V} \left.\frac{\Delta_2(\kappa R,\kappa r)}{\kappa^2}\right|_a^R =
\frac{\Delta_2(\kappa R,\kappa a)}{Z\lb},\qquad
\label{eqn:average1}\\ 
\left\langle \frac{\Delta_1^2(\kappa R,\kappa r)}{r^2}\right\rangle&=& 
\frac{4\pi}{V}\int_a^R\dd r\, \Delta_1^2(\kappa R,\kappa r) 
= -\frac{\kappa}{2Z\lb}\left\{\frac{\Delta_1(\kappa R,\kappa
    a)}{\kappa a} \times \right. 
\nonumber\\
&& \left.\times
\left[ \Delta_1(\kappa R,\kappa a)-\Delta_2(\kappa R,\kappa a)\right]-
4\kappa a \left(1-\kappa^2 R^2\right)-4\kappa^3 R^3
\vphantom{\frac{\Delta_1(\kappa R,\kappa a)}{\kappa a}} \right\}.
\label{eqn:average2}
\eea

Alternatively, Eq.~(\ref{eqn:dhpressuredirect}) may be checked by 
taking a \textit{formal} functional
derivative with respect to the WS-cell free volume $V$  
 --- analogously as performed in Appendix~\ref{app:a} for the 
nonlinear treatment  --- 
of the salt-free 
linearized Helmholtz free energy functional~(\ref{eqn:dhfunct}),
which is recast in the form 
\bea
\beta{\mathrsfs F}_\mathrm{DH}[n(\rb)]
&=& Z\left[\ln \left(\nc\zeta^3\right)-1\right]+
\int\dd^3\rb\, {f}[n(\rb)] , \\
{f}[n(\rb)]&=&\frac12\,\psi(\rb) \rho(\rb)
+\frac12\nc\left[\frac{n(\rb)}{\nc}-1\right]^2 ,
\eea
where $\rho(\rb)$ and $\psi(\rb)$ are defined 
in Section~\ref{sec:2} by 
Eqs.~(\ref{eqn:rho})~and~(\ref{eqn:pbpsi}), respectively.
The equilibrium counterion profile $\bar{n}(\rb)$ that minimizes
the linearized version of the extended Helmholtz free-energy 
functional ${\mathit{\tilde\Omega}}_\mathrm{DH}= {\mathrsfs F}_\mathrm{DH}- 
 \mu_\mathrm{el} \int\dd^3\rb \,n(\rb)$, with 
$\beta\bar\mu_\mathrm{el}=\left\langle\bar\psi\right\rangle$, is given 
by 
\bea
\bar{n}(\rb)&=& 
\nc \left[1+\left\langle\bar\psi\right\rangle -\bar{\psi}(\rb)\right].
\label{eqn:equilprofile}
\eea

As explained in Appendix~\ref{app:a} 
for the nonlinear PB treatment, it is much
easier to evaluate
\bea
\frac{\dd{\mathrsfs F}_\mathrm{DH}[\bar{n}(\rb)]}{\dd V }&=&
\left.\frac{\delta{\mathit {\tilde\Omega}}_\mathrm{DH}[{n}(\rb)]}
{\delta V}\right|_{\bar{n}(\rbs)} =
 \left.\frac{\partial {\mathit {\tilde\Omega}}_\mathrm{DH}[{n}(\rb)]}{\partial V }
\right|_{\bar{n}(\rbs)} +
\int\dd^3\rb'\left.\frac{\delta
{\mathit {\tilde\Omega}}_\mathrm{DH}[{n}(\rb)]}{\delta n(\rb')}\right|_{\bar{n}(\rbs')}
\frac{\dd\bar n(\rb')}{\dd V}, \qquad
\eea
because the second term vanishes due to the stationary condition. 
The first term reads
\bea
-\frac{\partial\beta{\mathit {\tilde\Omega}}_\mathrm{DH}[{n}(\rb)]}{\partial V }=
\nc\!&+&\!\beta\mu_\mathrm{el}
\,\frac{\partial}{\partial V}\int\dd^3\rb\,n(\rb)-
\frac{\partial}{\partial V}\int\dd^3\rb\, {f}[n(\rb)]-
\frac{\dd \nc}{\dd V}
\int\dd^3\rb\,\frac{\partial{f}[n(\rb)]}{\partial\nc},\qquad\\
\frac{\partial}{\partial V}
\int\dd^3\rb\, {f}[n(\rb)]
&=&\psi(R)n(R)
+\frac12\nc\left[\frac{n(R)}{\nc}-1\right]^2, \\
\frac{\dd \nc}{\dd V}
\int\dd^3\rb\,\frac{\partial{f}[n(\rb)]}{\partial\nc}&=&
-\frac{\nc}{V}\int\dd^3\rb\,\left\{
\frac12\left[\frac{n(\rb)}{\nc}-1\right]^2 -
\frac{n(\rb)}{\nc}\left[\frac{n(\rb)}{\nc}-1\right]\right\}.
\eea
Using the explicit form~(\ref{eqn:equilprofile}) 
of the linearized equilibrium counterion profile $\bar{n}(\rb)$, we obtain
\bea
\frac{\beta\bar\mu_\mathrm{el}}
{\nc}\left.\frac{\partial}{\partial V}\int\dd^3\rb\,n(\rb)
\right|_{\bar{n}(\rbs)}
&=& \left\langle\bar\psi\right\rangle
\left[1+\left\langle\bar\psi\right\rangle -\bar{\psi}(R)\right],\\
\left.\frac1{\nc}\,\frac{\partial}{\partial V}
\int\dd^3\rb\, {f}[n(\rb)]\right|_{\bar{n}(\rbs)}&=&
\bar\psi(R)\left[1+\left\langle\bar\psi\right\rangle -\bar{\psi}(R)\right]
+ \frac12 \left[\left\langle\bar\psi\right\rangle
  -\bar{\psi}(R)\right]^2,\\
\left.\frac1{\nc}\,\frac{\dd \nc}{\dd V}
\int\dd^3\rb\,\frac{\partial {f}[n(\rb)]}
{\partial\nc}\right|_{\bar{n}(\rbs)}&=&
-\frac12\left\langle\vphantom{\frac12}
\left[\left\langle\bar\psi\right\rangle-\bar\psi\right]^2\right\rangle
+ \left\langle\vphantom{\frac12}
\left[1+\left\langle\bar\psi\right\rangle-\bar\psi\right]
\left[\left\langle\bar\psi\right\rangle-\bar\psi\right]\right\rangle
\qquad\nonumber\\&=&
\frac12\left\langle\vphantom{\frac12}
\left[\left\langle\bar\psi\right\rangle-\bar\psi\right]^2\right\rangle,
\label{eqn:averagedelta2}
\eea 
which yields the salt-free linearized osmotic 
pressure Eq.~(\ref{eqn:dhpressuredirect}).

\section{Linearized osmotic pressure: canonical 
ensemble\label{app:d}}
\setcounter{equation}{0}

In this Appendix it will be shown that the linearized
canonical osmotic pressure is given by a quadratic 
expansion of the nonlinear canonical osmotic pressure.

The nonlinear PB Helmholtz free-energy functional,
\bea
\beta{\mathrsfs F}[n_\pm(\rb)]&=&
\frac1{8\pi\lb}\int\dd^3\rb\,\left[\nabla\psi(\rb)\right]^2
+ \sum_{i=\pm}\int\dd^3\rb\,n_i(\rb)\left\{\ln
  \left[{n_i(\rb)}\zeta^3_i\right]-1\right\} , \quad
\eea
is minimized with respect to $n_\pm(\rb)$ 
under the constraint of overall WS-cell
charge neutrality,~Eq.~(\ref{eqn:chargeneutralitysalt}).
We introduce a Lagrange multiplier $\mu_\mathrm{el}$
and define the extended Helmholtz free-energy functional, 
${\mathit{\tilde\Omega}}={\mathrsfs F}-
\mu_\mathrm{el}\int\dd^3\rb\, \rho(\rb)$.
Functional minimization of ${\mathit{\tilde\Omega}}$ with respect to
$n_\pm(\rb)$ yields the Boltzmann factors, ${n}_\pm(\rb)=
\mathrm{e}^{\pm\beta\mu_\mathrm{el}\mp\psi(\rbs)}/\zeta^3_\pm$.
The Lagrange multiplier,
\be 
\beta\mu_\mathrm{el}=
\ln\left(\bar n_+\zeta_+^3\right)-
\ln\left\langle\mathrm{e}^{-\psi({\rbs})}\right\rangle=
\ln\left(\bar n_-\zeta_-^3\right)+
\ln\left\langle\mathrm{e}^{\psi({\rbs})}\right\rangle,
\ee
is obtained by using the effective average definitions~(\ref{eqn:uniform}),
which automatically satisfy the  overall WS-cell
charge neutrality,~Eq.~(\ref{eqn:chargeneutralitysalt}).
Replacing into the Boltzmann factors, we obtain the nonlinear 
equilibrium density profiles, 
\be
n_\pm(\rb)= 
\frac{\bar{n}_\pm\,\mathrm{e}^{\mp\psi(\rbs)}}
{\left\langle \mathrm{e}^{\mp\psi(\rbs)}\right\rangle} =
\frac{\bar{n}_\pm
\exp\left[{\pm\left\langle\psi\right\rangle\mp\psi(\rb)}\right]}
{\left\langle\exp\left[{\pm\left\langle\psi\right\rangle\mp\psi(\rb)}
\right]\right\rangle}.
\ee
The simple ideal-gas relation~(\ref{eqn:bounddensity})
between the nonlinear osmotic pressure and the WS-cell
boundary density can be generalized in the presence of added salt.  
In this case, it can be shown that 
the nonlinear osmotic pressure is given by the total
microionic density, $n\equiv  {n}_++ n_-$, 
evaluated at the WS-cell boundary, 
\bea
\beta P &=& \left.-\frac{\dd \beta F}{\dd V}\right|_{Z,T,s}
= {n}(R)=
\frac{\bar{n}_+\,\mathrm{e}^{-\psi(R)}}
{\left\langle \mathrm{e}^{-\psi(\rbs)}\right\rangle}+ 
\frac{\bar{n}_-\,\mathrm{e}^{\psi(R)}}{\left\langle 
\mathrm{e}^{\psi(\rbs)}\right\rangle} . \label{eqn:pbcanonical}
\eea 

Let us now compare the linearized canonical osmotic 
pressure~(\ref{eqn:dhpressurecanonical}) with 
a quadratic expansion about the uniform
reference state of its nonlinear 
counterpart, Eq.~(\ref{eqn:pbcanonical}),
\bea
\beta P &=&
(1+s)\nc\frac{\exp\left[{\left\langle\psi\right\rangle-\psi(R)}\right]}
{\left\langle\exp\left[{\left\langle\psi\right\rangle-\psi(\rb)}\right]
\right\rangle}
+s\nc\frac{\exp\left[{-\left\langle\psi\right\rangle+\psi(R)}\right]}
{\left\langle\exp\left[{-\left\langle\psi\right\rangle+\psi(\rb)}\right]
\right\rangle}\nonumber\\
&=& (1+2s)\nc\left\{1+
\frac{1}{1+2s}\delta_1(R)+
\frac12\delta_2(R) 
-\frac12\left\langle\delta_2(\rb) \right\rangle
+{\mathrsfs O}\left[ \delta_3(R), 
\left\langle\delta_3(\rb) \right\rangle
\right]
\right\}, \label{eqn:pbpressurecanonical}\qquad
\eea
where the electrostatic potential differences $\delta_\nu(\rb)$
are given by~(\ref{eqn:deltan}).
Because of the redefinition of the screening length $\kappa^{-1}$
in terms of $\kc^{-1}$, 
the spatial averages~(\ref{eqn:average1})~and~(\ref{eqn:average2}) 
needed to evaluate $\left\langle\psi\right\rangle$ 
and $\left\langle\psi^2\right\rangle$ 
will be multiplied by a factor $(1+2s)^{-1}$. Using the explicit 
DH-like solution to the electrostatic potential~(\ref{eqn:dhsolution}),
it is indeed possible to show that the linearized canonical 
osmotic pressure~(\ref{eqn:dhpressurecanonical}) corresponds to the 
truncation of the expansion~(\ref{eqn:pbpressurecanonical}) up to
the quadratic terms,
i.e. $\beta P=\beta P_\mathrm{DH}^\mathrm{can}+
{\mathrsfs O}\left[ \delta_3(R), 
\left\langle\delta_3(\rb) \right\rangle\right]$,
with $\beta P_\mathrm{DH}^\mathrm{can}= n_\mathrm{DH}(R) + (1+2s)\,({\nc}/{2})
\left[ \delta_2(R)-\left\langle\delta_2(\rb) \right\rangle\right]$,
where the first term corresponds to the sum of the 
\textit{linearized} canonical densities at the WS-cell boundary.
We see that 
the linearized canonical osmotic pressure \textit{is not simply  given} by 
the linearized boundary density $n_\mathrm{DH}(R)$, 
because of the presence of the quadratic terms.

\section{Linearized osmotic pressure: 
semi-grand-canonical ensemble\label{app:e}}
\setcounter{equation}{0}

In this Appendix it will be shown that the linearized  
semi-grand-canonical osmotic pressure is given by a quadratic 
expansion of the nonlinear semi-grand-canonical osmotic pressure.

The nonlinear PB semi-grand-canonical functional associated to 
fixed microion chemical potentials
$\beta\mu_\pm=\ln \left(\nb \zeta^3_\pm\right)$,
\bea
\beta{\mathit{\Omega}}[n_\pm(\rb)] &\equiv& \beta{\mathrsfs F}[n_\pm(\rb)] 
-\sum_{i=\pm}\beta\mu_i\int\dd^3\rb\,n_i(\rb)
\nonumber\\&=&
\frac1{8\pi\lb}\int\dd^3\rb\,\left[\nabla\psi(\rb)\right]^2
+ \sum_{i=\pm}\int\dd^3\rb\,n_i(\rb)\left\{
\ln\left[\frac{n_i(\rb)}{\nb}\right]-1\right\} 
 ,\  \label{eqn:pbomegasphere}
\quad
\eea
is minimized with respect to $n_\pm(\rb)$ 
under the constraint of overall WS-cell
charge neutrality,~Eq.~(\ref{eqn:chargeneutralitysalt}).
We introduce a Lagrange multiplier $\mu_\mathrm{el}$
and define the extended semi-grand-canonical functional, 
${\mathit{\tilde\Omega}}={\mathit{\Omega}}-
\mu_\mathrm{el}\int\dd^3\rb\, \rho(\rb)$.
Functional minimization of ${\mathit{\tilde\Omega}}$ with respect to
$n_\pm(\rb)$ yields the Boltzmann factors, ${n}_\pm(\rb)=
\nb \mathrm{e}^{\pm\beta\mu_\mathrm{el}\mp\psi(\rbs)}$.
The Lagrange multiplier,
\be 
\mathrm{e}^{\pm\beta\mu_\mathrm{el}}=
 \frac{\sqrt{\nc^2 +
(2\nb)^2\left\langle 
\mathrm{e}^{\psi(\rbs)}\right\rangle\left\langle 
\mathrm{e}^{-\psi(\rbs)}\right\rangle}\pm \nc}{2\nb\left\langle 
\mathrm{e}^{\mp \psi(\rbs)}\right\rangle} ,
\ee
is obtained by imposing the overall WS-cell
charge neutrality,~Eq.~(\ref{eqn:chargeneutralitysalt}).
Replacing into the Boltzmann factors, we obtain the nonlinear 
equilibrium density profiles, 
\bea
n_\pm(\rb)&=& \frac{\sqrt{\nc^2 +
(2\nb)^2\alpha_+\alpha_-}\pm \nc}{2\alpha_\pm}
\, \mathrm{e}^{\pm\left\langle\psi\right\rangle \mp \psi(\rbs)}
= \left[\bar{n}_\pm +\frac14(1-\eta^2)\frac{\nc}{\eta}
\left\langle \delta_2(\rb)\right\rangle\right]\times\nonumber\qquad \\
&&\times \left[
1\pm\delta_1(\rb)
+\frac12 \delta_2(\rb) 
-\frac12\left\langle \delta_2(\rb)\right\rangle \right]
+{\mathrsfs O}\left[\delta_3(\rb),\left\langle \delta_3(\rb)\right\rangle
\right], \label{eqn:densitydonnan}\\
\alpha_\pm &=& \left\langle 
\mathrm{e}^{\pm \left\langle\psi\right\rangle \mp \psi(\rbs)}\right\rangle
= 1+\frac12\left\langle \delta_2(\rb)\right\rangle 
+{\mathrsfs O}\left[\left\langle \delta_3(\rb)\right\rangle
\right],\qquad
\eea
where the electrostatic potential differences $\delta_\nu(\rb)$
are given by~(\ref{eqn:deltan}).
At the nonlinear mean-field level the simple ideal-gas relation
between the osmotic pressure and the total  
microionic density, $n\equiv n_++n_-$,   
at the WS-cell boundary is still valid, 
leading to 
\bea
\beta P &=& \left.-\frac{\dd \beta\Omega}{\dd V}\right|_{Z,T,\mu_\pm}
= n(R)
=\frac{\sqrt{\nc^2 +
(2\nb)^2\alpha_+\alpha_-}+\nc}{2\alpha_+}
\, \mathrm{e}^{\left\langle\psi\right\rangle-\psi(R)}
+\frac{\sqrt{\nc^2 +
(2\nb)^2\alpha_+\alpha_-}-\nc}{2\alpha_-}
\, \mathrm{e}^{-\left\langle\psi\right\rangle+\psi(R)} 
\nonumber\\&=&
\frac{\nc}{\eta}
\left\{1+\eta\delta_1(R)
+\frac12\delta_2(R)
-\frac{\eta^2}2 \left\langle\delta_2(\rb)
\right\rangle
+{\mathrsfs O}\left[\delta_3(R),
\left\langle\delta_3(\rb)\right\rangle
\right]\right\}.\label{eqn:donnanpbpressure}
\eea 
Let us again compare the linearized semi-grand-canonical osmotic
pressure~(\ref{eqn:donnandhpressure}) 
with a quadratic expansion about the zero-th order Donnan
densities $\bar{n}_\pm$, Eqs.~(\ref{eqn:donnandensities}), 
of its nonlinear counterpart, Eq.~(\ref{eqn:donnanpbpressure}).
Now, because of the redefinition of $\kappa$, the 
spatial averages~(\ref{eqn:average1})~and~(\ref{eqn:average2}) 
needed to evaluate $\left\langle\psi\right\rangle$ 
and $\left\langle\psi^2\right\rangle$ are multiplied by 
a factor $\eta$.
Using again the explicit DH solution~(\ref{eqn:dhsolution}), 
it is possible to show that 
the linearized semi-grand-canonical 
osmotic pressure~(\ref{eqn:donnandhpressure}) corresponds to the
truncation of the expansion~(\ref{eqn:donnanpbpressure}) up to
the quadratic terms, i.e. 
$\beta P=\beta P_\mathrm{DH}^\mathrm{sgc} +{\mathrsfs O}\left[\delta_3(R),
\left\langle\delta_3(\rb)\right\rangle\right]$, 
with $\beta P_\mathrm{DH}^\mathrm{sgc}=n_\mathrm{DH}(R) +
[{\nc}/{(2\eta)}]\left[ \delta_2(R) -\eta^2  \left\langle\delta_2(\rb)
\right\rangle  \right]$, where the first term represents the sum 
of the \textit{linearized} semi-grand-canonical 
densities at the WS-cell boundary.
 We see again that the linearized semi-grand-canonical 
 osmotic pressure \textit{does not
correspond} to the linearized boundary density $n_\mathrm{DH}(R)$, 
because of the presence of the quadratic terms. 

\section{Linearized averaged densities for the 
semi-grand-canonical ensemble\label{app:f}}
\setcounter{equation}{0}

In this Appendix it will be shown that the self-consistent 
linearized averaged densities for the semi-grand-canonical
ensemble are given by the state-independent 
zero-th order Donnan densities,
which are obtained by a self-consistent minimization of the linearized 
semi-grand-canonical functional under the WS-cell charge-neutrality 
constraint.

To obtain the self-consistent averaged densities \textit{up to
second order} we need to expand the 
nonlinear semi-grand-canonical functional, Eq.~(\ref{eqn:pbomegasphere}), 
\textit{up to third order} 
about the (\textit{a priori} unknown) effective average densities 
$\left\langle n_\pm(\rb)\right\rangle\equiv 
(1/V)\int\dd^3\rb\,n_\pm(\rb)$, 
\bea
\beta{\mathit{\Omega}}[n_\pm(\rb)] 
&=&\frac1{8\pi\lb}\int\dd^3\rb\,\left[\nabla\psi(\rb)\right]^2
+V \sum_{i=\pm} \left\langle n_i\right\rangle
\left[\ln\frac{\left\langle n_i\right\rangle}{\nb} -1\right]
+ \sum_{i=\pm} \left[\left\langle n_i\right\rangle 
\ln \left\langle n_i\right\rangle \right] \int\dd^3\rb \,
\delta_i(\rb) \nonumber\\
&&+\frac12\sum_{i=\pm}\left\langle n_i\right\rangle 
\int\dd^3\rb\,\delta_i^2(\rb) 
-\frac16\sum_{i=\pm}\left\langle n_i\right\rangle 
\int\dd^3\rb \, \delta_i^3(\rb) 
+{\mathrsfs O} \left[\int\dd^3\rb\,\delta_i^4(\rb) \right],
\label{eqn:omegaexpansion}
\eea
where we introduced the relative deviations about the averaged
densities, 
\be
\delta_\pm(\rb) \equiv \frac{n_\pm(\rb)}{\left\langle n_\pm\right\rangle}-1. 
\ee
Keeping only quadratic terms in 
Eq.~(\ref{eqn:omegaexpansion}) is similar in spirit to the 
quadratic expansion presented in Eq.~(13) of 
Ref.~[\citen{deserno}]. However, we want to stress
that, in general, the unknown average expansion densities 
$\left\langle n_\pm\right\rangle$ depend itself 
on the ionic profiles. Functional minimization of 
Eq.~(\ref{eqn:omegaexpansion}) with respect to the profiles 
$n_\pm(\rb)$ must take this fact into account, in addition to the 
WS-cell charge-neutrality constraint, Eq.~(\ref{eqn:chargeneutralitysalt}). 
Eventually, for a linearized theory, the self-consistent 
expansion densities $\bar{n}_\pm$ turn out to be indeed  
independent on the ionic profiles, cf. Eq.~(\ref{eqn:donnandensities}),
but this can only be derived \textit{a posteriori.} 

We introduce a Lagrange multiplier $\mu_\mathrm{el}$ to ensure 
the overall WS-cell charge neutrality~(\ref{eqn:chargeneutralitysalt}),
and define the extended semi-grand-canonical functional, 
${\mathit{\tilde\Omega}}={\mathit{\Omega}}-
\mu_\mathrm{el}\int\dd^3\rb\, \rho(\rb)$.
Functional minimization of ${\mathit{\tilde\Omega}}$ with respect to
$n_\pm(\rb)$ leads to the Euler-Lagrange or stationary conditions,
${\delta{\mathit{\tilde\Omega}}}/{\delta n_\pm(\rb)}=0$, 
which may be cast in the form 
\be
n_\pm(\rb)= \left\langle n_\pm\right\rangle \left\{ 1
\mp\left[\psi(\rb)-\beta\mu_\mathrm{el}\right] - 
\ln\frac{\left\langle n_\pm\right\rangle}{\nb} 
+\frac12 \delta^2_\pm(\rb)+
{\mathrsfs O} \left[ \delta^3_\pm(\rb), 
\left\langle \delta^3_\pm(\rb) \right\rangle \right] 
\right\}, \label{eqn:npmquadratic}
\ee
where we have neglected the cubic averaged contribution, 
${\mathrsfs O} \left[\left\langle \delta^3_\pm(\rb) 
\right\rangle \right]$, 
because the neglected quartic term 
of~(\ref{eqn:omegaexpansion}) will also contributed to it. 
To obtain the averages $\left\langle n_\pm\right\rangle$ 
self-consistently, we integrate $n_\pm(\rb)$ over the 
volume to obtain the consistency relations,
\be
\left\langle n_\pm(\rb)\right\rangle =\nb \exp \left\{
\mp\left[\left\langle\psi(\rb)\right\rangle-\beta\mu_\mathrm{el}\right]
+\frac12 \left\langle \delta^2_\pm(\rb)\right\rangle +
{\mathrsfs O} \left[\left\langle \delta^3_\pm(\rb) \right\rangle
\right] \right\}, \label{eqn:avconsist}
\ee
where the Lagrange multiplier $\mu_\mathrm{el}$ is found 
by imposing the overall WS-cell charge 
neutrality~(\ref{eqn:chargeneutralitysalt}). 
We should stress that in addition to the chemical potential of 
microions $\mu$ that
defines the semi-grand-canonical ensemble, we introduced a
\textit{Lagrange multiplier} $\mu_\mathrm{el}$, whose 
role is twofold: besides the overall WS-cell charge 
neutrality~(\ref{eqn:chargeneutralitysalt}), it also ensures 
the gauge invariance of the electrostatic potential $\psi(\rb)$.
We should not confuse the chemical potential of microions 
$\mu$, which is fixed by the bulk salt concentration $\nb$ of the reservoir, 
with the linearized Lagrange multiplier $\mu_\mathrm{el}$,
which is associated with the Donnan effect and ensures the charge 
neutrality of the WS cell~(\ref{eqn:chargeneutralitysalt}). 

Before we derive 
the averages consistent with a \textit{linearization} of the 
PB equation, let us also obtain the self-consistent averages
corresponding to a \textit{quadratic approximation} of the 
nonlinear equation.
Noting that $\delta_\pm(\rb)=\pm\delta_1(\rb)+{\mathrsfs O} 
\left[ \delta_2(\rb), 
\left\langle \delta_2(\rb) \right\rangle \right]$
and neglecting cubic terms 
in~Eqs.~(\ref{eqn:npmquadratic})~and~(\ref{eqn:avconsist})
leads to the quadratic self-consistent averages,
\be
\left\langle n_\pm (\rb) \right\rangle = \frac{\sqrt{\nc^2 +
(2\nb)^2 
\mathrm{e}^{\left\langle \delta_2(\rbs)\right\rangle} }\pm\nc}{2},
\label{eqn:donnanquadratic}
\ee 
and the quadratic equilibrium density profiles,
\be
n_\pm(\rb)=\left\langle n_\pm(\rb)\right\rangle \left[ 1
\pm \delta_1(\rb) 
+\frac12 \delta_2(\rb) - 
\frac12\left\langle \delta_2(\rb) \right\rangle \right],
\ee
where the electrostatic potential $\nu$-th order difference 
$\delta_\nu(\rb)$ is defined by~(\ref{eqn:deltan}). 
These correspond indeed to the \textit{quadratic} expansions
of the nonlinear PB average densities and equilibrium density profiles,
respectively, and are correct up to $\delta_2(\rb)$ (quadratic) 
terms. Here we may see another 
advantage of the gauge-invariant formulation: it provides us a 
systematic way to consider self-consistent higher-order approximations
of the nonlinear equations, while the fixed-gauge analysis of 
Deserno and von Gr\"unberg \cite{deserno} does not allow them to 
extend their calculations to include higher-order terms.

However, in order to be consistent with a \textit{linearization} of the 
PB equation one needs to neglect also the quadratic terms in the
approximate Euler-Lagrange conditions~(\ref{eqn:npmquadratic}), 
although \textit{global} self-consistency of the 
Legendre transformation will require to include them --- 
 cf. Appendix~\ref{app:g}.
It is clear that ignoring these terms will 
yield the \textit{state-independent} 
zero-th order Donnan densities as the self-consistent 
linearized averaged densities, 
\be
\left\langle n_\pm(\rb)  \right\rangle = \frac{\sqrt{\nc^2 +
(2\nb)^2}\pm\nc}{2},
\ee 
and the linearized equilibrium density profiles,
\be
n_\pm(\rb)=\left\langle n_\pm(\rb)\right\rangle 
\left[ 1\pm \delta_1(\rb)\right]. \label{eqn:npmlinear}
\ee
In the main text we used the notation $\bar{n}_\pm \equiv 
\left\langle n_\pm  \right\rangle_1$, Eq.~(\ref{eqn:donnandensities}), 
to refer to the 
linearized self-consistent averaged densities for the 
semi-grand-canonical ensemble, where the subscript `1' emphasizes
the fact that the average densities were obtained under linearization. 
Deserno and von Gr\"unberg~\cite{deserno} justify this 
choice for the expansion densities --- 
written in terms of an optimal linearization point $\bar\psi_\mathrm{opt}$
defined by $\bar{n}_\pm=\nb\mathrm{e}^{\mp\bar\psi_\mathrm{opt}}$ ---  
by arguing that any other choice for the linearization point 
would lead to conflicting inequalities involving 
nonlinear and linearized averages. 
In a gauge-invariant formulation, however, the justification 
is indeed based on the self-consistent \textit{minimization} of the linearized 
semi-grand-canonical functional 
${\mathit{\Omega}}_\mathrm{DH}[n_\pm(\rb)]$, which is obtained 
by truncating the expansion of the 
nonlinear functional ${\mathit{\Omega}}[n_\pm(\rb)]$,
given by Eq.~(\ref{eqn:omegaexpansion}), only 
up to the quadratic terms and neglecting 
(consistently under linearization) the quadratic contribution 
in the approximated 
averaged Euler-Lagrange equations, Eq.~(\ref{eqn:avconsist}). 
Although \textit{internal} self-consistency 
(within the semi-grand-canonical ensemble) is achieved by using 
the linearized self-consistent averaged 
densities~(\ref{eqn:donnandensities}), we show in 
Appendix~\ref{app:g} that 
\textit{global} self-consistency under linearization  
(between the canonical and the semi-grand-canonical ensembles)
will also require the inclusion of the quadratic state-dependent 
terms of the self-consistent averaged 
densities~(\ref{eqn:donnanquadratic}). 

\section{Legendre transformation at the 
linearized level\label{app:g}}
\setcounter{equation}{0}

In this Appendix we discuss the 
differences between the Legendre transformation 
connecting the
canonical and the semi-grand-canonical ensembles 
at the level of the linearized functionals (before 
the functional minimization) and of the linearized 
thermodynamic potentials (after the functional 
minimization). It is shown that, in order to preserve 
the exactness of the Legendre transformation, quadratic 
contributions to the linearized expansion densities 
should be included in the former case, which are 
automatically included in the latter case.

At the nonlinear PB level the osmotic pressures obtained in 
the two distinct (canonical and semi-grand-canonical) ensembles are
completely equivalent \cite{footnote8}, provided we map them using the 
nonlinear relation
\be
(\nc+\ns)\ns= 
\left\langle n_+(\rb)\right\rangle \left\langle n_-(\rb)\right\rangle
= \nb^2 \left\langle \mathrm{e}^{\psi(\rbs)}\right\rangle \left\langle
\mathrm{e}^{-\psi(\rbs)}\right\rangle, \label{eqn:nonlinearmap}
\ee
where $\ns\equiv\left\langle n_-(\rb)\right\rangle$ is the effective average 
salt concentration in the colloidal suspension. 
The exact (at the mean-field level) relation (\ref{eqn:nonlinearmap}) 
follows directly from the gauge-invariant forms of the nonlinear 
average densities~(\ref{eqn:densitydonnanaverage}). 
Therefore, up to quadratic order, the linearized osmotic pressures 
are related by the renormalization of the
total average density of microions in the two ensembles,
\be
n\equiv\nc+2\ns=\left(1+2s\right)\nc\to
\left\langle n_+(\rb)\right\rangle+\left\langle n_-(\rb)\right\rangle=
\frac{\nc}{\eta}\left\{1+\frac12\left(1-\eta^2\right) 
\left\langle\delta_2(\rb)\right\rangle
+{\mathrsfs O}\left[\left\langle\delta_3(\rb)
\right\rangle\right]\right\}. \!\!\!
\label{eqn:legendre}
\ee
Because of the quadratic contribution,
the linearized osmotic pressures obtained in the two ensembles 
\textit{do not have the same form} when they are mapped 
using the zero-th order renormalization $(1+2s)\to {\eta}^{-1}$.
In other words, although for the nonlinear equations 
the Legendre transformation between the 
canonical and the semi-grand-canonical ensembles is exact, 
the same does not hold for the linearized equations:
one needs to use the approximated mapping~(\ref{eqn:legendre}) 
and expand consistently the linearized osmotic pressure up to 
quadratic-order terms.
This introduces an additional source of deviations for the
 linearized  semi-grand-canonical equations of state. 
In particular, the thermodynamically-conjugated density  
(in the semi-grand-canonical ensemble) to the 
chemical potential of salt particles, $\mus=\mu_++\mu_-$,
with $\mu_\pm= \beta^{-1}\ln\left(\nb\zeta_\pm^3\right)$, 
that corresponds to the (effective) 
total average density of microions inside the colloidal 
suspension, $n=\bar{n}_++\bar{n}_-$, 
\bea
n&\equiv& -\frac{2\dd}{\dd\mus}
\left[\frac{\Omega_\mathrm{DH}}{V}\right]_{\np}=
\frac{\nc}{\eta}
\left\{1+\frac{\eta^2}2\left(\eta^2-1\right)
+ \frac{Z\kappa\lb\eta}{4\Delta_2^2(\kappa R,\kappa a)}
\left(\eta^2-1\right) \times
\right.\nonumber\\
&&\left.\times\left[\frac{\Delta_1(\kappa R,\kappa a)}{\kappa a}
\left[ \Delta_1(\kappa R,\kappa a)-\Delta_2(\kappa
R,\kappa a)\right] -4\kappa a \left(1-\kappa^2 R^2\right)-4\kappa^3 R^3 
\vphantom{\frac{\eta^2}{2}}\right] \vphantom{\frac{\eta^2}{2}}\right\},
\eea
is indeed given by the right-hand side 
of~Eq.~(\ref{eqn:legendre}) neglecting cubic and 
higher-order contributions. This conjugated density, 
however, is \textit{inconsistent} (up to the quadratic order,
but consistent under \textit{linearization}) with the 
state-independent zero-th order Donnan densities~(\ref{eqn:donnandensities}), 
i.e. $n\neq \bar{n}_++\bar{n}_-=\nc/\eta$, because of the presence 
of the quadratic
contribution in Eq.~(\ref{eqn:legendre}).

An alternative procedure to the Legendre transformation 
of the linearized Helmholtz free-energy \textit{functional} 
${\mathrsfs F}_\mathrm{DH}[n_\pm(\rb)]$ --- which is done, 
as presented in Appendix~\ref{app:f}, \textit{before} the functional
minimization with respect to the profiles --- 
is to perform it \textit{after} the functional minimization, 
at the Helmholtz \textit{free-energy} level. 
Because in the canonical ensemble the expansion densities 
$\bar{n}_\pm$ are known \textit{a priori}, the Legendre 
transformation that maps the linearized Helmholtz free energy 
${F}_\mathrm{DH}$ into the linearized semi-grand-canonical potential 
$\Omega_\mathrm{DH}$ can then be obtained without any further 
approximations for the expansion densities. 
For this purpose it is convenient to recall the definition of the 
\textit{total} volume of 
the WS cell and of the \textit{nominal} densities of 
counterions, polyions and salt particles, 
\be
\tilde{V}\equiv \frac{V}{1-\phi},
\qquad \tilde\nc\equiv\nc\left(1-\phi\right),
\qquad \np\equiv \frac{1}{\tilde{V}}
= \phi \left(\frac{4\pi}{3} a^3\right)^{-1},
\qquad \tilde\ns\equiv\ns\left(1-\phi\right). 
\ee
It is important to stress that the \textit{nominal} densities
should be used, instead of the \textit{effective} ones.
Introducing the linearized Helmholtz free-energy density, 
$f_\mathrm{DH}\equiv F_\mathrm{DH}/\tilde{V}$,
where $F_\mathrm{DH}$ in the presence of added salt 
is given by~Eq.~(\ref{eqn:freesalted}), 
one may check that the Legendre transformation at the 
linearized free-energy level is indeed \textit{exact,} 
since the linearized chemical potentials 
of salt particles and of polyions,
\bea
\beta\mus&\equiv& 
\left.\frac{\dd \beta f_\mathrm{DH}}{\dd \tilde\ns}\right|_{\np}
= \ln \left[(1+s)\nc\zeta^3_+\right] 
+ \ln\left(s\nc\zeta^3_-\right) +
\frac{1}{\left(1+2s\right)^2}+ \frac{1}{1+2s}\, 
\frac{Z\kappa\lb}{2\Delta_2^2(\kappa R,\kappa a)}\times\nonumber\\
&&\times\left[\frac{\Delta_1(\kappa R,\kappa a)}{\kappa a}
\left[ \Delta_1(\kappa R,\kappa a)-\Delta_2(\kappa R,\kappa a)\right]-
4\kappa a \left(1-\kappa^2 R^2\right)-4\kappa^3 R^3 \right],
\label{eqn:musalt} \\
\beta\mup&\equiv& 
\left.\frac{\dd \beta f_\mathrm{DH}}{\dd \np}\right|_{\tilde\ns}
= \np^{-1}\left\{\tilde\nc\ln \left[(1+s)\nc\zeta^3_+\right]
+\frac{\tilde\nc}2  \left[\frac{Z\lb}{a}\, 
\frac{\Delta_1(\kappa R,\kappa a)}{\Delta_2(\kappa R,\kappa a)} - 
\frac1{1+2s}\right] -\frac{s\tilde\nc}{\left(1+2s\right)^2} 
\right.\nonumber\\
&& +(1+2s)\nc
\left[\phi+\frac{2Z^2\kappa^2\lb^2}{\Delta_2^2(\kappa R,\kappa a)}\right] 
+ \frac{\left(1+2s\phi\right)\nc}
{1+2s}\,\frac{Z\kappa\lb}{4\Delta_2^2(\kappa R,\kappa a)} \times \nonumber\\
&&\left.\times \left[\frac{\Delta_1(\kappa R,\kappa a)}{\kappa a}
\left[ \Delta_1(\kappa R,\kappa a)-\Delta_2(\kappa R,\kappa a)\right]-
4\kappa a \left(1-\kappa^2 R^2\right)-4\kappa^3 R^3 \right]\right\},
\label{eqn:mumacroion} 
\eea
satisfy the thermodynamical identity 
\be
P_\mathrm{DH}^\mathrm{can} = \tilde\ns \mus + \np \mup
- f_\mathrm{DH}, \label{eqn:gibbs}
\ee
where the linearized \textit{canonical} osmotic pressure 
entering into~Eq.~(\ref{eqn:gibbs}),
$P_\mathrm{DH}^\mathrm{can}$, 
is given by Eq.~(\ref{eqn:dhpressurecanonical}). 
It should also be remarked that Eq.~(\ref{eqn:musalt}) 
corresponds to the truncation of the expansion of the 
exact nonlinear mapping~(\ref{eqn:nonlinearmap})
up to quadratic-order terms,
\be
\beta\mus = \ln\left[\frac{(\nc+\ns)\ns \zeta^3_+ \zeta^3_-}
{\left\langle \mathrm{e}^{\psi(\rbs)}\right\rangle \left\langle
\mathrm{e}^{-\psi(\rbs)}\right\rangle}\right] =
\ln \left[(1+s)\nc\zeta^3_+\right] 
+ \ln\left(s\nc\zeta^3_-\right) -
\left\langle \delta_2(\rb)\right\rangle
+{\mathrsfs O} \left[\left\langle \delta_3(\rb)\right\rangle \right].
\ee

The thermodynamical relation~(\ref{eqn:gibbs}) can also be viewed as 
defining the Legendre transformation.
Instead of obtaining the osmotic-pressure isotherms for a constant 
number of salt particles inside the WS cell (canonical case,
fixed $s$), we may consider them at fixed chemical 
potential of salt particles $\mus$  
 (semi-grand-canonical case), 
which corresponds to a system in electrochemical equilibrium 
with an infinite salt 
reservoir of bulk density $\nb$, defined by 
\be
\beta\mus\equiv \beta\mu_++\beta\mu_-=  
\ln \left(\nb^2\zeta^3_+\zeta^3_-\right) . 
\label{eqn:snb}
\ee
Solution of the nonlinear equation defined 
by~(\ref{eqn:musalt}) and (\ref{eqn:snb}) yields 
the Legendre transformation in an implicit parametric form, 
\be
s(\phi,\nb)=\frac{\sqrt{1+
\left[2\nb/\nc(\phi)\right]^2 
\mathrm{e}^{\left\langle \delta_2(\rbs)\right\rangle}}-1}{2}, 
\qquad\quad\nc(\phi)= \frac{3Z}{4\pi a^3}\left(\frac{\phi}{1-\phi}\right),  
\ee
with the quadratic electrostatic-potential deviation in the 
canonical ensemble given by 
\bea
\left\langle \delta_2(\rb)\right\rangle &=&
-\frac{1}{\left(1+2s\right)^2} - \frac{1}{1+2s}\, 
\frac{Z\kappa\lb}{2\Delta_2^2(\kappa R,\kappa a)}\times\nonumber\\
&&\times
\left[\frac{\Delta_1(\kappa R,\kappa a)}{\kappa a}
\left[ \Delta_1(\kappa R,\kappa a)-\Delta_2(\kappa R,\kappa a)\right]-
4\kappa a \left(1-\kappa^2 R^2\right)-4\kappa^3 R^3 \right].\qquad
\label{eqn:averaged2}
\eea
The linearized \textit{semi-grand-canonical} osmotic pressure, 
$\hat{P}_\mathrm{DH}^\mathrm{sgc}=
\hat{P}_\mathrm{DH}^\mathrm{sgc}(\phi,\nb)$, is then obtained 
by inserting the Legendre transformation $s=s(\phi,\nb)$ 
into the linearized \textit{canonical} osmotic pressure 
$P_\mathrm{DH}^\mathrm{can}=P_\mathrm{DH}^\mathrm{can}(\phi,s)$, 
Eq.~(\ref{eqn:dhpressurecanonical}). In other words, the two 
linearized osmotic pressures are related by 
$\hat{P}_\mathrm{DH}^\mathrm{sgc}(\phi,\nb)\equiv 
P_\mathrm{DH}^\mathrm{can}\left[\phi,s(\phi,\nb)\right]$. 
We should remark that the linearized semi-grand-canonical
osmotic pressure obtained by this 
procedure, $\hat{P}_\mathrm{DH}^\mathrm{sgc}(\phi,\nb)$, 
\textit{does not} coincide with 
its counterpart $P_\mathrm{DH}^\mathrm{sgc}(\phi,\nb)$ given by  
Eq.~(\ref{eqn:donnandhpressure}).

The disagreement between the two distinct linearized 
semi-grand-canonical osmotic pressures, 
$\hat{P}_\mathrm{DH}^\mathrm{sgc}(\phi,\nb)\neq
{P}_\mathrm{DH}^\mathrm{sgc}(\phi,\nb)$ --- 
obtained by Legendre transformations performed  
pre- and pos-minimization of the 
linearized functional 
${\mathrsfs F}_\mathrm{DH}[n_\pm(\rb)]$  
with respect to the profiles  --- 
may be traced back to the neglected quadratic 
contribution in the self-consistent linearized averaged 
densities~(\ref{eqn:donnandensities}).
Although the neglect of this state-dependent contribution in 
the average densities~(\ref{eqn:donnanquadratic}) is
justified to ensure mathematical consistency of the truncation
under the linearized approximation,
in order to obtain \textit{global} thermodynamic 
self-consistency (i.e., in order to preserve the \textit{exactness}) 
of the Legendre transformation  
one needs to keep \textit{all terms} of the quadratic truncation  
of the Euler-Lagrange equations, regardless of its inconsistency 
from the mathematical point of view. 
This leads to the average 
densities~(\ref{eqn:donnanquadratic}), which include
the quadratic state-dependent 
contribution $\left\langle\delta_2(\rb)\right\rangle$
 ---  in contrast to~Eqs.~(\ref{eqn:donnandensities}), 
which neglect it. 
Compared to the linearized semi-grand-canonical potential 
 $\Omega_\mathrm{DH}$, 
Eq.~(\ref{eqn:omegadh}), which uses the state-independent zero-th order 
Donnan densities~(\ref{eqn:donnandensities}) as 
expansion densities,
an augmented version $\hat\Omega_\mathrm{DH}$ 
using the quadratic average densities~(\ref{eqn:donnanquadratic}) 
will have an additional quadratic state-dependent contribution, 
\bea
\beta \hat\Omega_\mathrm{DH}&=&Z \left[\arctanh\hat\eta-
\frac1{\hat\eta}- \frac{\hat\eta}{2}
+ \frac{Z\lb}{2a}\, 
\frac{\Delta_1(\hat\kappa R,\hat\kappa a)}{\Delta_2(\hat\kappa R,
\hat\kappa a)} 
+\frac{1}{2\hat\eta} 
\left\langle \delta_2(\rb)\right\rangle \right],
\eea
where the parameter $\hat\eta$ and the (effective) Debye screening length 
$\hat\kappa^{-1}$ in the suspension, 
\be
\hat\eta\equiv\frac{\nc}{\sqrt{\nc^2+(2\nb)^2
\mathrm{e}^{\left\langle \delta_2(\rbs)\right\rangle}}}, 
\qquad\qquad \hat\kappa^2 =\frac{\kappa_\mathrm{c}^2}{\hat\eta} = 
\frac{\kb^2 \mathrm{e}^{\left\langle \delta_2(\rbs)\right\rangle/2}}
{\sqrt{1-\hat\eta^2}},\qquad \label{eqn:etaquadratic}
\ee
are now given implicitly in terms of the 
quadratic electrostatic-potential 
deviation in the \textit{semi-grand-canonical}
ensemble, which is obtained by replacing $(1+2s)\to \hat\eta^{-1}$ 
and $\kappa\to\hat\kappa$ 
in the expression of its canonical counterpart, Eq.~(\ref{eqn:averaged2}). 
These yield the \textit{globally} self-consistent (i.e., that
preserve the exactness of the Legendre transformation)
linearized semi-grand-canonical equations of state,  
\bea
\hat{n}&\equiv& -\frac{2\dd}{\dd\mus}
\left[\frac{\hat\Omega_\mathrm{DH}}{V}\right]_{\np} =
\frac{\nc}{\hat\eta} ={\sqrt{\nc^2+(2\nb)^2
\mathrm{e}^{\left\langle \delta_2(\rbs)\right\rangle}}}=
 \left\langle{n}_+(\rb)\right\rangle+
\left\langle{n}_-(\rb)\right\rangle, \label{eqn:nlegendre}\\
\beta\hat P_\mathrm{DH}^\mathrm{sgc} &\equiv&
-\left.\frac{\dd\beta\hat\Omega_\mathrm{DH} }{\dd V}\right|_{Z,T,\mu_\pm}=
\frac{\nc}{\hat\eta}\left\{1
+ \frac{Z\hat\kappa\lb\hat\eta}{4\Delta_2^2(\hat\kappa R,\hat\kappa
  a)}\left[\frac{\Delta_1(\hat\kappa R,\hat\kappa a)}{\hat\kappa a}
\,\times\right.\right. \nonumber \\
&&\times\left.\left.
\left[ \Delta_1(\hat\kappa R,\hat\kappa a)-\Delta_2(\hat\kappa R,
\hat\kappa a)\right]-
4\hat\kappa a \left(1+\frac23\hat\kappa^2 a^2-\hat\kappa^2 R^2\right)-
\frac43\hat\kappa^3 R^3 
\right]\right\}, \qquad\label{eqn:dhsemigrandpressure}
\eea
where the total derivatives must take the 
$\left\langle \delta_2(\rb)\right\rangle$ parametric 
implicit dependence of Eqs. (\ref{eqn:etaquadratic}) into account. 
The (effective) total average density of microions $\hat{n}$, 
Eq.~(\ref{eqn:nlegendre}),
the linearized semi-grand-canonical 
osmotic pressure $\hat{P}_\mathrm{DH}^\mathrm{sgc}$, 
Eq.~(\ref{eqn:dhsemigrandpressure}), and the 
linearized chemical potential of polyions 
$\mup(\phi,\nb)$
are now fully consistent with their canonical counterparts, 
given respectively by 
$n=(1+2s)\nc$ and Eqs.~(\ref{eqn:dhpressurecanonical})
and~(\ref{eqn:mumacroion}). 
They are related by 
the Legendre transformation
$(1+2s) =\hat\eta^{-1}$, where $\hat\eta$ --- given implicitly 
by~Eq.~(\ref{eqn:etaquadratic}) --- also includes 
quadratic state-dependent contributions.
A further Legendre transformation of the linearized 
semi-grand-canonical potential $\hat\Omega_\mathrm{DH}$ 
regains, as it should, 
the linearized pressure $\hat{P}_\mathrm{DH}^\mathrm{sgc}$, 
\be
\hat{P}_\mathrm{DH}^\mathrm{sgc} = \np \mup - 
\frac{\hat\Omega_\mathrm{DH}}{\tilde{V}}.
\ee

\begin{figure}[hp]
\begin{center}
\epsfig{file=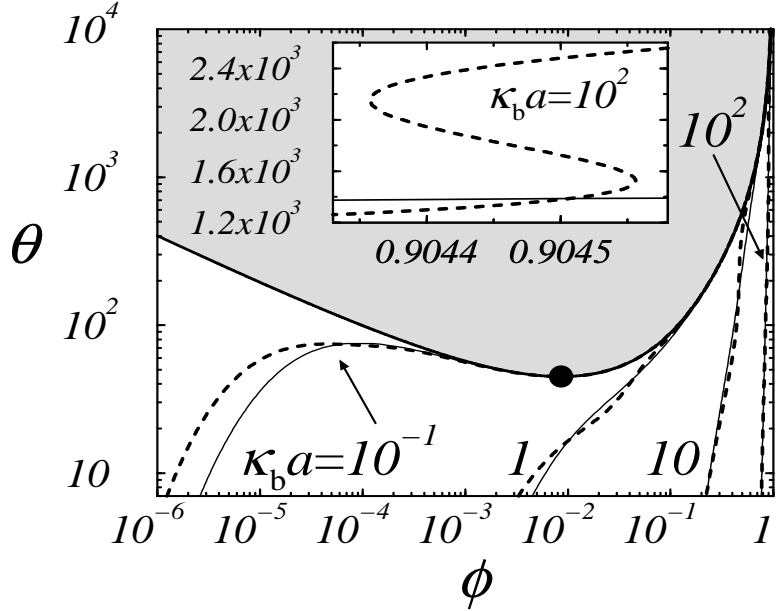,width=0.65\textwidth}
\caption{Spinodal lines $(\chi_\mathrm{DH}^{-1}=0)$
associated with the linearized semi-grand-canonical
osmotic pressures $P_\mathrm{DH}^\mathrm{sgc}$ (dashed lines), 
Eq.~(\ref{eqn:donnandhpressure}), and 
$\hat{P}_\mathrm{DH}^\mathrm{sgc}$ (solid lines), 
Eq.~(\ref{eqn:dhsemigrandpressure}), 
in the $\theta=3Z\lb/a$ versus 
volume fraction $\phi=(a/R)^3$ plane. 
They delimit the spurious 
unstable region that extends to lower values of $\phi$.
To allow a comparison with the canonical case (Figure~\ref{fig:2}), 
we also show the salt-free critical point (black circle) and 
the salt-free $(\kb a=0)$ unstable gray region. 
In the salt-free limit $(\kb a\ll 1)$ the semi-grand-canonical 
spinodal line reduces to the salt-free one, although for 
any nonvanishing $\kb a$ eventually 
it  will bend to the zero-temperature critical 
point at $(\phi_\mathrm{crit},\theta_\mathrm{crit})=(0,0)$.
Contrary to the canonical case, an increase of the bulk
reservoir density in the semi-grand-canonical case 
 \textit{enhances the instability,} as can be seen
from the different spinodal lines with increasing $\kb a$.
A typical monotonic (non-oscillating) 
osmotic-pressure isotherm is presented in Figure~3 (dotted curve) 
in Ref.~[\citen{deserno}]. See also main text for
additional comments.\label{fig:3}}
\end{center}
\end{figure}

The \textit{globally} self-consistent  
linearized osmotic pressure  $\hat{P}_\mathrm{DH}^\mathrm{sgc}$, 
 Eq.~(\ref{eqn:dhsemigrandpressure}), 
which preserves the exactness of the Legendre transformation, 
leads to results qualitatively similar to those of 
Eq.~(\ref{eqn:donnandhpressure}). In particular, 
the nonmonotonic behaviour of the spinodal lines for 
weak screening $(\kb a\ll 1)$ and the intrinsic instability of the 
low-$\phi$ phase are still predicted by $\hat{P}_\mathrm{DH}^\mathrm{sgc}$, 
as shown in Figure~\ref{fig:3}, where we compare the spinodal
lines associated with the two 
distinct semi-grand-canonical linearized osmotic-pressure definitions, given 
by~Eqs.~(\ref{eqn:donnandhpressure})~and~(\ref{eqn:dhsemigrandpressure}).
We should mention, however, that \textit{explicit analytical} comparison
in the exactly solvable planar case \cite{plane} does not show 
any improvement of the agreement between the nonlinear and 
linearized equations with the 
inclusion of the quadratic contribution to the average densities. 
Any \textit{numerical} indications in this 
direction, which were indeed observed in the planar case \cite{plane}, 
 are purely fortuitous. 
In fact, asymptotic expansions in the weak-coupling $(\lb\to 0)$ 
and in the ideal-gas limit of
both linearized osmotic pressures in the planar case, 
${P}_\mathrm{DH}^\mathrm{sgc}$ and
$\hat{P}_\mathrm{DH}^\mathrm{sgc}$, 
agree with the full nonlinear PB version up to 
the \textit{same order}.

\begin{thebibliography}{99}
\bibitem{footnote3}
Rigorously one
should take the total derivative with respect to the \textit{total} 
WS-cell volume, 
$\tilde{V}\equiv 1/\np=V/(1-\phi)$. However, because we are 
assuming that the polyions are 
incompressible, $\dd\tilde{V}=\dd V$.
\bibitem{footnote7}
Note that the nonlinear osmotic-pressure difference 
satisfies $\beta\Delta P\equiv n_+(R)+n_-(R)-2\nb=
4\nb \sinh^2\left\{\frac12\left[\beta\mu_\mathrm{el}-
\psi(R)\right]\right\}>0$.
\end{thebibliography}
\end{document}